\documentclass[a4paper,11pt]{article}
\usepackage{jheppub}

\usepackage[american]{babel}
\usepackage{amsfonts}
\usepackage{amsmath}
\usepackage{amssymb}
\usepackage{slashed}
\usepackage{xcolor}
\usepackage{bm}
\usepackage{float}
\usepackage[utf8]{inputenc}
\usepackage[macrosonly]{chet}
\usepackage{nicefrac}
\usepackage{hyperref}
\usepackage{url}
\usepackage{array}
\usepackage{multirow}
\usepackage{comment}
\usepackage{nicefrac}
\usepackage[normalem]{ulem}

\usepackage[makeroom]{cancel}
\usepackage{listings}

\definecolor{mmkeyword}{RGB}{0, 0, 200}
\definecolor{mmstring}{RGB}{180, 0, 0}
\definecolor{mmcomment}{RGB}{100, 100, 100}
\definecolor{mmbackground}{RGB}{248, 248, 248}

\lstdefinelanguage{Mathematica}{
  keywords={Series, Total, Flatten, First, True, False},
  keywordstyle=\color{mmkeyword}\bfseries,
  sensitive=true,
  comment=[l]{(*},
  morecomment=[s]{(*}{*)},
  commentstyle=\color{mmcomment}\itshape,
  stringstyle=\color{mmstring},
  morestring=[b]",
}

\usepackage{mathtools,tensor,mhchem}

\usepackage{lmodern}
\usepackage{tikz}
\usetikzlibrary{decorations.markings}
\usetikzlibrary{decorations.pathmorphing}
\usetikzlibrary{positioning}

\tikzset{graviton/.style={decorate, decoration={snake, segment length=2mm, amplitude=0.6mm}}}

\tikzset{
    mid arrow/.style={
        thick,
        postaction={
            decorate,
            decoration={
                markings,
                mark=at position #1 with {\arrow{>}}
            }
        }
    }
}

\abstract{
Irreversibility theorems -- such as the $A$-theorem -- establish a hierarchy among fixed points of the renormalization group flow.
The strongest thesis of this type of theorems would be that there exists a scalar function $A$ (generally suggested by the topological Weyl anomaly) and a positive definite metric $G_{IJ}$ in the space of couplings such that the renormalization group flow satisfies a gradient equation, $\partial_I A= G_{IJ}\beta^J$, in which case $A$ is locally monotonic along the flow.
In this paper we consider the long-range multiscalar $\phi^4$ theory, a theory without a local energy-momentum tensor that is unitary in $d=2,3$ and that is believed to be conformally invariant at fixed points, and show that its renormalization group flow satisfies the gradient structure up to the third loop order in the coupling.
We also show that $A$ and $G_{IJ}$ can be matched to the leading nontrivial order with the sphere free-energy $\tilde{F}$ and Zamolodchikov's metric $C_{IJ}$ of the corresponding conformal theory concentrating on the examples of the long-range vector $O(N)$ and hypercubic $H_N$ models.
Our results imply a perturbative proof of the $\tilde{F}$-theorem at the leading nontrivial order. We conclude the paper discussing briefly whether this result should hold to the next orders in perturbation theory.  

}

\begin{document}

\title{Matching $\bm{A}$ with $\bm{F}$ in long-range QFTs}

\author{Lorenzo Benfatto, }
\emailAdd{lorenzo.benfatto@phd.unipi.it}
\affiliation{Universit\`a di Pisa and INFN - Sezione di Pisa, Largo Bruno Pontecorvo 3, 56127 Pisa, Italy}

\author{Omar Zanusso}
\emailAdd{omar.zanusso@unipi.it}

\maketitle

\section{Introduction}\label{sect:intro}

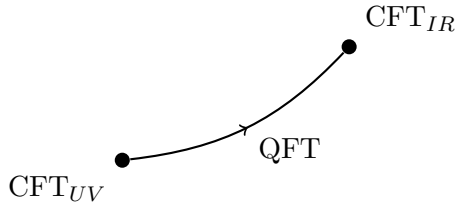
\begin{figure}[t]
\centering
\begin{tikzpicture}[
    dot/.style={circle, fill, inner sep=2pt},
    arrow/.style={->, thick},
]
\def\xA{0}
\def\yA{0}
\def\xB{3}
\def\yB{1.5}
\def\bend{-20} 
\node[dot, label=below left:{${\rm CFT}_{UV}$}] (A) at (\xA,\yA) {};
\node[dot, label=above right:{${\rm CFT}_{IR}$}] (B) at (\xB,\yB) {};
\draw[mid arrow=0.5] (A) to[bend left=\bend] node[midway, below right] {${\rm QFT}$} (B);
\end{tikzpicture}
\caption{Interpretation of a QFT as the RG trajectory connecting ${\rm CFT}_{UV}$ and ${\rm CFT}_{IR}$. With some tweaks this interpretation can be made to accommodate also effective field theories.}
\label{fig:CFT-QFT}
\end{figure}

A popular and formal point of view is to understand some quantum field theories (QFTs)
as whatever happens in the intermediate renormalization group (RG) flow between two fixed points which may be described by conformal field theories (CFTs), as depicted in Fig.~\ref{fig:CFT-QFT}. Since the RG trajectory flows from the ultraviolet (UV) to the infrared (IR), it establishes a relative hierarchy between ${\rm CFT}_{UV}$ and ${\rm CFT}_{IR}$.
From this point of view the natural question to ask is whether this hierarchy is universal, i.e., it does not depend on the specific trajectory that connects the two CFTs and the RG flow is irreversible. From a Wilsonian perspective the existence of a hierarchy is natural because the RG represents a coarse-graining of the degrees of freedom, which naturally flows only one way.

There are several theorems, which we broadly call \textit{monotonicity theorems}, that establish irreversibility in the space of field theories -- either locally or globally -- under reasonable sets of assumptions. A nonexhaustive list includes the famous $c$-theorem in $d=2$ \cite{Zamolodchikov:1986gt, Cappelli:1990yc, Shore:2016xor, Osborn:1991gm, Casini:2006es}, the $F$-theorem in $d=3$ \cite{Klebanov:2011gs, Casini:2012ei}, the $A$-theorems in $d=4$ \cite{Cardy:1988cwa, Osborn:1989td, Jack:1990eb, Komargodski:2011vj, Komargodski:2011xv, Luty:2012ww}, the $g$-theorem for boundaries and defects \cite{Jensen:2015swa, Kobayashi:2018lil, Casini:2018nym}, the $k$-theorem for charged degrees of freedom \cite{Nakamura:2025hyw} and higher dimensional generalizations \cite{Elvang:2012st, Grinstein:2014xba, Cordova:2015fha, Stergiou:2016uqq}.

One strength of monotonicity theorems is that they help in charting the landscape of possible CFTs often inducing a natural classification based on the degrees of freedom of any given RG fixed point and in doing so they may rule out flows that do not describe an averaging process over the QFTs degrees of freedom.
The ``downside'' is that the applicability of the theorems is tied to the assumptions made in their proof. Standard assumptions generally include unitarity (or reflection-positivity in the Euclidean case) and the presence of a local energy-momentum tensor (EMT).\footnote{%
These assumptions are not universal, for example unitarity could be relaxed in favor of $\mathcal{PT}$-invariance~\cite{Castro-Alvaredo:2017udm,Diatlyk:2026oxm}.}
However, it is not difficult to find examples of QFTs to which the above assumptions do not apply, but we still expect irreversibility. For example, models that would be unitary CFTs in integer dimension become nonunitary in noninteger dimensions \cite{Hogervorst:2014rta, Hogervorst:2015akt}, but they still seem to satisfy nontrivial monotonicity constraints \cite{Pannell:2024sia}.

With the above considerations in mind, we choose to consider the long-range $\phi^4$ theory and its generalizations, which are field theories that lack a local EMT and can be defined in arbitrary dimension $d$, but are also argued to be described by CFTs at fixed points.\footnote{%
As we discuss in the next section, Ref.~\cite{Paulos:2015jfa} argues that the long-range Ising model is described by a CFT at the critical point that is captured by the long-range $\phi^4$ field theory with one scalar, but a similar argument can be extended to the general long-range $\phi^4$ with arbitrary number of scalar flavors.}
Long-range theories have sparked interest since the seminal work of Fisher \cite{Fisher:1972zz} and describe the universality classes of lattice models of spins
interacting with a strength that is inversely proportional to some power of the distance between them. 
These models have received particular attention in the past years, both in the general case and in the particular examples regarding long-range interactions between vector spins \cite{Defenu:2014bea, Chai:2021arp, Behan:2023ile, Benedetti:2024wgx, Fraser-Taliente:2025udk, Benedetti:2020rrq, Giombi:2019enr, Behan:2025ydd}.

In this paper we first investigate whether the RG flow of the multiscalar long-range $\phi^4$ field theory admits a gradient structure pertubatively, meaning that there exists a monotonic function $A$ whose gradient gives the beta functions $\beta^I$ up to a positive-definite metric $G_{IJ}$ in the space of couplings, i.e.\ 
\begin{equation}
 \partial_I A = G_{IJ} \beta^J\,,
\end{equation}
which is one of the strongest forms of monotonicity theorem and, as such, the one that constrains the landscape the most. We find that, given the known result for $\beta^I$ to the third nontrivial order in a perturbative expansion, a parametric family of solutions for $A$ and $G_{IJ}$ always exists.

Then we compute directly the CFT candidates for the monotonic function and the metric tensor at the fixed points in conformal perturbation theory.
In particular, we derive the analitically continued sphere free-energy $\tilde{F}$ \cite{Giombi:2014xxa, Fei:2015oha, Giombi:2015haa} and (the coefficient of) the renormalized two-point function of quasi-marginal operators $C_{IJ}$, which generalizes Zamolodchikov's metric on theory space.
Importantly, we show that it is possible to match the RG results with the CFT results, which implies that
\begin{equation}\label{eq:Gradient-Flow}
    \partial_I \tilde{F}=C_{IJ}\beta^J \, ,
\end{equation}
holds at the leading nontrivial order in perturbation theory.
This procedure fixes the free parameters that are left undetermined by the RG procedure and gives a nontrivial confirmation of the $F$-theorem in the case of long-range models to the leading order. Whether this is true to all orders in perturbation theory is up to debate and we return to this point in the conclusions.

The paper is structured as follows.
Section \ref{sect:phi4LR} presents a brief overview of the theoretical framework of the multiscalar long-range $\phi^4$ theory with some additional context for our work.
Section \ref{sect:consistency} includes the solution to the gradient flow equations in the long-range theory up to the third order in perturbation theory.
Section \ref{sect:F-function} is devoted to the computation of the renormalized sphere free-energy, then specialized to the (vector) $O(N)$ and (hypercubic) $H_N$ fixed points to obtain the $\tilde{F}$-function, while in Section \ref{sect:two-point} we use the same methods to derive an expression for the generalization of Zamolodchikov's metric $C_{IJ}$. 
In Section \ref{sect:matching} we show the matching of RG and CFT results. In Section \ref{sect:conclusions} we conclude by discussing whether we should expect this result to hold at the next orders in perturbation theory.

\section{From the long-range Ising model to the long-range $\phi^4$ theory}\label{sect:phi4LR}

The long-range (LR) lattice Ising model is described by the nonlocal Hamiltonian
\begin{equation}\label{eq:lattice-lr-ising}
 {\cal H} = - \sum_{ab} J_{ab} \sigma_a \sigma_b \,, \qquad J_{ab} = \frac{J}{r_{ab}^{d+s}}\,,
\end{equation}
where the summation extends over the sites of a $d$-dimensional lattice with spins $\sigma_a=\pm 1$
and $r_{ab}$ is the distance between any two given sites. The parameter $s$ is free and changes the strength of the LR interaction. If $s$ is sufficiently large and positive we expect that the interaction becomes effectively short-range (SR), implying that a genuinely long-range universality class should interpolate with the one of the standard short-range Ising model for some value of $s=s^*(d)$ as discussed below.

We now introduce a generalization of the nonlocal continuum field theory that is believed to represent the universality class of \eqref{eq:lattice-lr-ising}.
We define the LR multiscalar quartic field theory in general dimension $d$ through the flat-space action
\begin{equation}\label{eq:action}
    S[\phi]=\int {\rm d}^d x \left[ \frac{1}{2}  \phi_i(x) ( - \partial^2 )^{\frac{s}{2}} \phi_i(x)+\frac{1}{4!} \lambda_{ijkl} \, \phi_i(x) \phi_j(x) \phi_k(x) \phi_l(x) \right] \, ,
\end{equation}
where the flavor scalar indices span $i=1,\cdots,N$, and the coupling $\lambda_{ijkl}$ is a fully symmetric tensor.
The universality class of the LR Ising model corresponds to the critical point of the case $N=1$, in which case the tensor $\lambda_{ijkl}$ reduces to a single coupling.\footnote{%
The euclidean LR $\phi^4$ theory, as well as several generalizations, have been shown to be reflection positive in $d=3$ \cite{Gubinelli:2021nou}.
}

The relation with \eqref{eq:lattice-lr-ising} is immediately evident if the fractional power of the Laplacian is expressed in terms of the Euclidean distance as
\begin{equation}\label{eq:free-action}
  \int {\rm d}^d x   \frac{1}{2}  \phi_i(x) (-\partial^2)^{\frac{s}{2}} \phi_i(x)
 = \frac{2^{s-1}\Gamma\left(\frac{d+s}{2}\right)}{\pi^{d/2}\Gamma\left( -\frac{s}{2} \right)} \int {\rm d}^d x\, {\rm d}^d y \, \phi_i(x) \frac{1}{|x-y|^{2(d-\Delta_\phi)}}\phi_i(y)\,,
\end{equation}
where $\Delta_\phi=\frac{d-s}{2}$ is the dimension of the scalar operator. Eq.~\eqref{eq:free-action} could be regarded as the definition of the fractional Laplacian. The propagator in coordinate space is
\begin{equation}\label{eq:propagator-flat}
  G_{d,s}(x,y)
 \, 
 =  \frac{C_{d,s}}{|x-y|^{d-s}} \,, \qquad C_{d,s}=\frac{\Gamma(\frac{d-s}{2})}{\pi^{\frac{d}{2}}2^s\Gamma(\frac{s}{2})} \, ,
\end{equation}
and it is diagonal in field space.
Our choice of normalization for the two-point function matches that in \cite{Giombi:2022gjj}.
Naively, for $s>0$ the interaction is long-range except for $s=2$, in which case it coincides with the standard $\phi^4$ theory.

Despite the many similarities with the standard short-range field theory, the RG analysis of \eqref{eq:action} is unique and interesting for several reasons, including the fact that critical models may (and will) exist as functions of $s$.
In fact, for arbitrary $s$ the dimension of the fields $\phi_i$ is fixed to be $[\phi_i]=\frac{d-s}{2}$, implying that the fields $\phi_i$ do not have an anomalous dimension, so the bare and renormalized fields coincide. This is consistent with the fact that there are no counterterms to Eq.~\eqref{eq:free-action}, i.e., fractional powers of the momentum are not renormalized in typical dimensionless subtraction schemes.

Dimensional analysis shows that the $\phi^4$ interaction is marginal for $s=\frac{d}{2}$.
It is thus possible to introduce a parameter $\varepsilon$, \emph{distinct} from the one of dimensional regularization, and parametrize $s=\frac{d+\varepsilon}{2}$, which ensures that $[\phi^4]=d-\varepsilon$ and the interaction term contains relevant operators when $\varepsilon>0$.
For $s<\frac{d}{2}$, i.e.\ $\varepsilon<0$, the interaction is irrelevant and the RG is governed by the Gaussian fixed point. Instead for $s>\frac{d}{2}$, i.e.\ $\varepsilon>0$ the interaction is a relevant deformation and the RG is governed by a nontrivial fixed point, see for example Ref.~\cite{Benedetti:2020rrq} and references therein.

\begin{figure}[ht]
\centering
\includegraphics[scale=1]{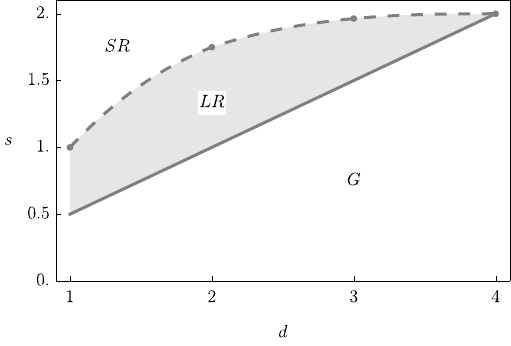}
\caption{The three different phases of the critical point in the long range Ising model. The dashed line is the curve $s_*(d)=2-\eta_{SR}(d)$, interpolated by with the known behaviors in the integer dimensions $d=1,2,3$ and the perturbative expansion below $d=4$.}
\label{fig:LRI}
\end{figure}

Ref.~\cite{Paulos:2015jfa} discusses the phase diagram of the LR/SR Ising models in the $(d,s)$ plane. We depict it in Fig.~\ref{fig:LRI}, which displays all possible phases: $G$ indicates the Gaussian behavior, $SR$ and $LR$ the short- and long-range ones.
The solid line is the $s=\frac{d}{2}$ line separating $\varepsilon<0$ below and $\varepsilon>0$ above.
The dashed line parametrizes $s^*(d)=2-\eta_{SR}(d)$, where $\eta_{SR}$ is the anomalous dimension of the field for the standard short-range Ising model. Starting from the solid line at fixed $d$, for increasing $s$ we have that at some point $s=s^*(d)$ the dimension $\Delta_\phi=(d-s)/2$ intercepts the value $\Delta_{SR}= (d-2+\eta_{SR})/2$ of the SR model and above that the universality class coincides with the standard SR one. In short: the above arguments suggest that at any fixed $d<4$ there is a nontrivial phase for $\frac{d}{2}<s <s^*(d)$ which is genuinely long-range, while for $s>s^*(d)$ the critical behavior coincides with the one of the short-range model.
At the point $s=s_*$ there is evidence of a nonperturbative crossover between the short and the long-range theories \cite{Fisher:1972zz, Angelini:2014koy, Brezin:2014rkt}, which, however, may also admit a weakly coupled description \cite{Behan:2017dwr, Behan:2017emf}.

The physics we wish to discuss in this paper lies in the lower part of the LR region of Fig.~\ref{fig:LRI}, where it is possible to study the field theory perturbatively in $\varepsilon$. From now on we thus restrict ourselves to the neighborhood of the $s=\frac{d}{2}$ line, working in $d=2$ and $d=3$ for small $\varepsilon>0$.
Evidence given in Ref.~\cite{Paulos:2015jfa} points to the field theory of the long-range Ising model as being conformally invariant at the critical point. However, based on the same arguments given above, a diagram equivalent to Fig.~\ref{fig:LRI} could be drawn for nonlocal theories in which the potential $\frac{1}{4!}\lambda_{ijkl} \, \phi_i\phi_j\phi_k \phi_l$ enjoys other symmetries for $N>1$ (examples being the $O(N)$, hypercubic and all other symmetries of a multiscalar $\phi^4$), which is why we consider the most general case.

Operatively, we compute Feynman diagrams within the loop expansion and renormalize the model subtracting $\frac{1}{\varepsilon}$ poles at fixed $d$ (recall that here $\varepsilon$ parametrizes $s$, so it is not the usual small parameter of dimensional regularization).
Starting from there, we assume that conformal invariance holds at all fixed points of the long-range $\phi^4$ theory as well, and study the presence of a gradient structure in terms of a function $A$ and a metric $G_{IJ}$ along the RG flow.
By considering the fixed point to be described by CFTs, we then match $\tilde{F}$ (the free energy on a sphere up to a constant) with $A$ and $G_{IJ}$ with the coefficient $C_{IJ}$ of a two-point function that is defined in complete analogy with Zamolodchikov's original paper.

\section{Structure of the expansion and consistency conditions}\label{sect:consistency}

We denote the tensor coupling of the action \eqref{eq:action} as $\lambda^I=\lambda_{ijkl}$, where $I=\{ ijkl \}$ is a collective index spanning over totally symmetric combinations of the scalar flavor indices. For the scope of this section we rescale the couplings of the action \eqref{eq:action} as
\begin{equation}\label{eq:rescaling-benedetti}
 \lambda^I \to \alpha \, \lambda^I \qquad {\rm with} \qquad \alpha=(4 \pi)^{d/2} \Gamma(d/2)
\end{equation}
in order to match our beta functions coefficients with those in \cite{Benedetti:2020rrq}. The couplings source the operators $O_I(x)=\phi_i\phi_j\phi_k\phi_l$, so that in this notation the potential is simply $V(\phi)=\lambda^I O_I$.
Ultraviolet divergences appear as poles $\frac{1}{\varepsilon}$ in the loop expansion, that are minimally subtracted through counterterms in the usual way. As discussed in the previous section, the dimension $d$ can be kept fixed and there are only counterterms to the vertex and not to the wavefunction in the long-range case. The renormalization procedure results in a nontrivial tensor beta function, denoted $\beta^I$ and expressed as a series in contractions of $\lambda^I$ with appropriate coefficients.

Up to $L'$ loops the beta function has the form
\begin{equation}
 \beta^I = -\varepsilon \lambda^I + \sum_{L=1}^{L'} \beta^{(L), I}\,, \qquad \beta^{(L)} = \sum_\alpha  b_{L,\alpha}(\varepsilon) \left(\lambda\right)_\alpha^{L+1}
\end{equation}
where $L\leq L'$ is any given loop order and $\alpha$ spans over all possible one-vertex irreducible contractions of $L+1$ copies of $\lambda^I$ with four external legs that are schematically represented as $\left(\lambda\right)_\alpha^{L+1}$ (tensor indices are suppressed in the second formula to simplify the notation).
The general beta function, which we use in this section, is known up to $L'=3$ \cite{Benedetti:2020rrq}.

The peculiarity of the long-range model is that the coefficients of the terms of the beta function, $b_{L,\alpha}(\varepsilon)$, are themselves expressed as series in $\varepsilon$, in agreement with the fact that the universality class exists parametrically as a function of $s=\frac{d+\varepsilon}{2}$. 
In general the coefficients $b_{L,\alpha}(\varepsilon)$ are determined to loop order $L'$ as the finite series
\begin{equation}
 b_{L,\alpha}(\varepsilon) = \sum_{p=0}^{L'-L} b^{(p)}_{L,\alpha} \varepsilon^p \,.
\end{equation}
The above structure is consistent with the fact that $\lambda \sim \varepsilon$ at fixed points, so at any loop order $L'$ the entire correction of order $\varepsilon^{L'+1}$ of $\beta$ is determined,
and hence fixed points and critical exponents are known to order $\varepsilon^{L'}$.

In this section we verify the gradient structure to the RG flow, i.e., we construct
\begin{equation}\label{eq:gradient}
    \partial_I A=G_{IJ}\beta^J \ ,
\end{equation}
where $A$ is a scalar function of the tensor couplings, $\partial_I=\partial/\partial \lambda^I$ is the derivative with respect to the couplings,
and $G_{IJ}$ may be interpreted as a metric on the theory space.
The structure \eqref{eq:gradient} is guaranteed for the RG flow of local CFTs
as demonstrated using the local RG approach and the Weyl consistency conditions of the conformal anomaly in even $d$,
in which case $A$ is identified with the topological Weyl anomaly.
General applications of this approach
are in Refs.~\cite{Osborn:1991gm, Jack:2013sha, Poole:2018ljd, Grinstein:2013cka, Nakayama:2013wda, Jack:2014pua, Poole:2019kcm}, while the original idea can be traced back to \cite{Wallace:1974dx}. It is straightforward to prove that $A$ is a monotonic function along the RG if the metric is positive definite because $\mu \frac{d}{d\mu} A = \beta^I \partial_I A = G_{IJ}\beta^I\beta^J \geq 0$.

Requiring that Eq.~\eqref{eq:gradient} is valid perturbatively means obtaining a set of equations that relate order-by-order the coefficients of $A$ and $G$ with those of $\beta$. This set of equations can then be solved to express the coefficients of $A$ and $G$ as functions of those of $\beta$. In our case the solution is not unique up to a set of undetermined parameters.
It is worth noting that the existence of a solution is not guaranteed in general at higher loops. For example, one can eliminate the coefficients of $A$ and $G$ from the system of equations to find a set of constraints that the coefficient of $\beta$ should satisfy. Those can then be checked against explicit computation of the $\beta$ to the given loop order.
For example, this approach has been applied in Ref.~\cite{Pannell:2024sia} for the multiscalar SR $\phi^4$ theory, in Ref.~\cite{Gracey:2015fia, Benfatto:2025yci} for the multiscalar SR $\phi^3$ and in Ref.~\cite{Pannell:2025ajf} for SR scalar-fermion QFTs.
However, nontrivial consistency constraints coming from the integrability of Eq.~\eqref{eq:gradient} generally arise at higher loop order (five loops for the case of SR $\phi^4$) and, in fact, the beta function of the action \eqref{eq:action} will be integrable to order $L'=3$.

In the next few subsections we consider equation \eqref{eq:gradient} in an expansion in orders of $\lambda$ starting from $\lambda$, which we refer to as $0$-loop, up to $\lambda^4$,  which we refer to as $3$-loop. This nomenclature is based on the number of loops of the diagrams that represent the contractions of the coupling tensor (that are not Feynman diagrams, but rather the result of the computation of a Feynman diagram).

\subsection{$0$-loop consistency}\label{sect:0loop}

%

We represent everything diagrammatically: the couplings are dotted vertices and contractions between them are solid lines. In front of each diagram we include a symmetry factor that represents the possible equivalent permutations.
At this order, the contribution to the beta function is the trivial scaling part that is proportional to $\varepsilon$:
\begin{equation}\label{eq:beta0}
  \beta^{(0)}_{ijkl} = -\varepsilon \, {\cal S}_1\,
  \begin{tikzpicture}[scale=0.5,baseline=(vert_cent.base)]
	\node (vert_cent) at (0,0) {$\phantom{\cdot}$};
 \node (center) at (0,0) {};
	\def\radius{1cm};
	    \node[inner sep=0pt] (tl) at (150:1.5\radius) {};
        \node[inner sep=0pt] (bl) at (210:1.5\radius) {};
        \node[inner sep=0pt] (tr) at (30:1.5\radius) {};
        \node[inner sep=0pt] (br) at (330:1.5\radius) {};
	    \filldraw (center) circle[radius=2pt];
	\draw (tl) -- (br);
	\draw (bl) -- (tr);
  \end{tikzpicture} \ .
\end{equation}
Also at this order the metric is trivial, i.e., $G^{(0)}_{IJ} \propto \delta_{IJ}$ and its leading coefficient can be chosen at will, corresponding to an arbitrary normalization. We choose to fix its leading $\varepsilon^0$ coefficient to $m_0^{(0)}=1$, so that the metric becomes
\begin{equation}\label{eq:metric0}
 G^{(0)}_{IJ} = m_0 \
 \,{\cal S}_{24}\,\left[ \,
    \begin{tikzpicture}[scale=0.5,baseline=(vert_cent.base)]
	\node (vert_cent) at (0,0) {$\phantom{\cdot}$};
	\node (center) at (0,0) {};
	\def\vlength{0.7cm};
        \def\hlength{1cm};
        \node[inner sep=0pt] (l1) at (-1*\hlength,1.25*\vlength) {};
        \node[inner sep=0pt] (l2) at (-1*\hlength,0.5*\vlength) {};
	\node[inner sep=0pt] (l3) at (-1*\hlength,-0.25*\vlength) {};
 	\node[inner sep=0pt] (l4) at (-1*\hlength,-1*\vlength) {};
	\node[inner sep=0pt] (r1) at (1*\hlength,1.25*\vlength) {};
	\node[inner sep=0pt] (r2) at (1*\hlength,0.5*\vlength){};
        \node[inner sep=0pt] (r3) at (1*\hlength,-0.25*\vlength) {};
        \node[inner sep=0pt] (r4) at (1*\hlength,-1*\vlength) {};
        \draw (l1) -- (r1);
        \draw (l2) -- (r2);
        \draw (l3) -- (r3);
        \draw (l4) -- (r4);
    \end{tikzpicture}
   \, \right] \ , \qquad m_0 = 1 + \varepsilon \ m_0^{(1)} + \varepsilon^2 \ m_0^{(2)} \,.
\end{equation}
There is only one contraction of the tensor coupling that is available at this order for the $A$-function
\begin{equation}\label{eq:afun0}
 \begin{split}
 A^{(0)} &= 
 a_0\,
  \begin{tikzpicture}[scale=0.5,baseline=(vert_cent.base)]
	\node (vert_cent) at (0,0) {$\phantom{\cdot}$};
	\node (center) at (0,0) {};
	\def\radius{1cm};
	\node[inner sep=0pt] (l) at (0:\radius) {};
	\node[inner sep=0pt] (r) at (180:\radius) {};
	\draw (center) circle[radius=\radius];
	\filldraw (r) circle[radius=2pt];
    \filldraw (l) circle[radius=2pt];
	\draw (r) to[out=45, in=135] (l);
 \draw (r) to[out=-45, in=-135] (l);
	\end{tikzpicture}
 \ , \qquad a_0 = a_0^{(0)}+\varepsilon a_0^{(1)} + \varepsilon^2 \ a_0^{(2)}  + \varepsilon^3 \ a_0^{(3)} \, ,
 \end{split}
\end{equation}
Now we use Eq.~\eqref{eq:gradient} to order $\lambda$
with $\beta^{(0)}$, to determine linear relations among the coefficients of the terms of $G$ and $A$ given above. The linear relations are matched order-by-order in $\varepsilon$ up to order $L'-L=3$. The general result is determined up to two free parameters,
\begin{equation}\label{eq:consistency-0loop}
        a_0^{(0)}= 0
        \,,\quad
        a_0^{(1)}= -\frac{1}{2}
        \,,\quad
        a_0^{(2)}= -\frac{\alpha_0}{2}
        \,,\quad
        a_0^{(3)}= -\frac{\alpha_1}{2}
        \, ,
\end{equation}
where we have chosen $\alpha_0=m_0^{(1)}$ and $\alpha_1=m_0^{(2)}$. Here
and in the following we always choose to express components of the metric, rather than of $A$, as free parameters.

\subsection{$1$-loop consistency}\label{sect:1loop}

For the next order we proceed in analogy with the previous one. 
The beta function is
\begin{equation}\label{eq:beta1}
  \beta^{(1)}_{ijkl} = b_{1}\,{\cal S}_3\,
  \begin{tikzpicture}[scale=0.5,baseline=(vert_cent.base)]
	\node (vert_cent) at (0,0) {$\phantom{\cdot}$};
 \node (center) at (0,0) {};
	\def\radius{1cm};
	\node[inner sep=0pt] (l) at (180:\radius) {};
        \node[inner sep=0pt] (etl) at (160:1.7*\radius) {};
        \node[inner sep=0pt] (ebl) at (200:1.7*\radius) {};
        \node[inner sep=0pt] (itr) at (20:1.7*\radius) {};
        \node[inner sep=0pt] (ibr) at (-20:1.7*\radius) {};
        \node[inner sep=0pt] (r) at (0:\radius) {};
	       \draw (center) circle[radius=\radius];
	       \filldraw (l) circle[radius=2pt];
        \filldraw (r) circle[radius=2pt];
	\draw (l) -- (etl);
	\draw (l) -- (ebl);
	\draw (r) -- (itr);
    \draw (r) -- (ibr);
  \end{tikzpicture}
\, , \qquad
    b_1 = b_1^{(0)}+\varepsilon \, b_1^{(1)} + \varepsilon^2 \, b_1^{(2)} \ ,
\end{equation}
and the coefficients can be found in Ref.~\cite{Benedetti:2020rrq}
\begin{equation}\label{eq:beta1-values}
 b_1^{(0)}= 1
 \,, \quad 
 b_1^{(1)}= \frac{1}{2}\left(\psi(1)-\psi(d/2)\right)
 \,, \quad 
 b_1^{(2)}= \frac{1}{8}\left\{
 \left(\psi(1)-\psi(d/2)\right)^2 + \psi_1(1)-\psi_1(d/2)
 \right \}\,,
\end{equation}
where $\psi(z)=\ln\Gamma(z)$ and $\psi_m(z)= \frac{{\rm d}^m}{{\rm d}z^m}\ln \Gamma(z)$.
The diagram representing the contraction in \eqref{eq:beta1} justifies why we refer to $1$-loop at this order, but we remark that some of the coefficients in Eq.~\eqref{eq:beta1-values} actually require computations of Feynman diagrams up to three loops.
The next contribution to the metric can be parametrized as
\begin{equation}\label{eq:metric1}
 G^{(1)}_{IJ} =
 m_{1}\,{\cal S}_{72}\,\left[ \,
   \begin{tikzpicture}[scale=0.5,baseline=(vert_cent.base)]
	\node (vert_cent) at (0,-0.5) {$\phantom{\cdot}$};
 \node (center) at (0,0) {};
	\def\radius{0.7cm};
 	\def\vlength{0.5cm};
        \def\hlength{1cm};
	\node[inner sep=0pt] (tl) at (135:0) {};
        \node[inner sep=0pt] (etl) at (160:1.7*\radius) {};
        \node[inner sep=0pt] (bl) at (225:0) {};
        \node[inner sep=0pt] (ebl) at (200:1.7*\radius) {};
        \node[inner sep=0pt] (tr) at (20:1.7*\radius) {};
        \node[inner sep=0pt] (br) at (-20:1.7*\radius) {};
        \node[inner sep=0pt] (itr) at (45:0) {};
        \node[inner sep=0pt] (ibr) at (-45:0) {};
        \node[inner sep=0pt] (l3) at (-1.*\hlength,-1.5*\vlength) {};
 	      \node[inner sep=0pt] (l4) at (-1*\hlength,-2.5*\vlength) {};
        \node[inner sep=0pt] (r3) at (1*\hlength,-1.5*\vlength) {};
        \node[inner sep=0pt] (r4) at (1*\hlength,-2.5*\vlength) {};
	       \filldraw (tl) circle[radius=2pt];
        \filldraw (bl) circle[radius=2pt];
        \filldraw (itr) circle[radius=2pt];
        \filldraw (ibr) circle[radius=2pt];
	\draw (tl) -- (etl);
	\draw (bl) -- (ebl);
	\draw (itr) -- (tr);
    \draw (ibr) -- (br);
    \draw (l3) -- (r3);
    \draw (l4) -- (r4);
  \end{tikzpicture}
   \, \right]
\, , \qquad
    m_1 = m_1^{(0)}+\varepsilon \, m_1^{(1)} \, .
\end{equation}
We note in passing that \eqref{eq:metric1} is the first nontrivial metric contribution and the computation of its analog in conformal perturbation theory from the two-point function $\langle \phi_I \phi_J\rangle$ is the subject of Section \ref{sect:two-point}.
The next term of the $A$-function is parametrized as
\begin{equation}\label{eq:afun1}
 \begin{split}
 A^{(1)} &= 
 a_1\,
  \begin{tikzpicture}[scale=0.5,baseline=(vert_cent.base)]
	\node (vert_cent) at (0,0) {$\phantom{\cdot}$};
	\node (center) at (0,0) {};
	\def\radius{1cm};
	\node[inner sep=0pt] (tr) at (0:\radius) {};
	\node[inner sep=0pt] (tl) at (120:\radius) {};
	\node[inner sep=0pt] (bl) at (240:\radius) {};
	\draw (center) circle[radius=\radius];
	\filldraw (tr) circle[radius=2pt];
        \filldraw (tl) circle[radius=2pt];
        \filldraw (bl) circle[radius=2pt];
	\draw (tr) to[out=180, in=-60] (tl);
	\draw (tl) to[out=-60, in=60] (bl);
 \draw (tr) to[out=180, in=60] (bl);
	\end{tikzpicture}
 \end{split}
 \,, \qquad
    a_1 = a_1^{(0)}+\varepsilon \, a_1^{(1)} + \varepsilon^2 \, a_1^{(2)} \ .
\end{equation}
The solution of \eqref{eq:gradient} to order $\lambda^2$ gives
\begin{equation}\label{eq:consistency-1loop}
     a_1^{(0)}= b_1^{(0)}
     \,, \quad
     a_1^{(1)}= -\frac{\alpha_2}{3}+\alpha_0 b_1^{(0)}+b_1^{(1)} 
     \,, \quad
     a_1^{(2)}= -\frac{\alpha_3}{3}+\alpha_1 b_1^{(0)}+\alpha_0 b_1^{(1)}+b_1^{(2)} \, ,
\end{equation}
where we introduced two additional free parameters, $\alpha_2=m_1^{(0)}$ and $\alpha_3=m_1^{(1)}$. The final result can be obtained inserting the values given in Eq.~\eqref{eq:beta1-values} in the above relations. Notice that a solution is always possible independently of the specific coefficients of the beta function, and the same applies to the next two orders given below.
Moreover this is the last order for which the solution is self-contained within this loop order and the previous one, meaning that all the $a$s can be determined from the $b$s and the $m$s that do not come from the solutions at higher orders.

\subsection{$2$-loop consistency}\label{sect:2loop}

At this order the beta functions of the LR and SR model differ ``structurally'' because in the LR case there are no terms from the gamma-functions (bare and renormalized fields coincide in the long-range case).
The contribution to the beta function is thus a one-vertex irreducible contraction
\begin{equation}\label{eq:beta2}
 \begin{split}
  \beta^{(2)}_{ijkl} &=
  b_{2}\,{\cal S}_6\,
  \begin{tikzpicture}[scale=0.5,baseline=(vert_cent.base)]
	\node (vert_cent) at (0,0) {$\phantom{\cdot}$};
	\node (center) at (0,0) {};
	\def\radius{1cm};
	\node[inner sep=0pt] (tl) at (120:\radius) {};
        \node[inner sep=0pt] (etl) at (120:1.7*\radius) {};
        \node[inner sep=0pt] (bl) at (240:\radius) {};
        \node[inner sep=0pt] (ebl) at (240:1.7*\radius) {};
	\node[inner sep=0pt] (r) at (0:\radius) {};
        \node[inner sep=0pt] (etr) at (30:1.7*\radius) {};
        \node[inner sep=0pt] (ebr) at (-30:1.7*\radius) {};
	\draw (center) circle[radius=\radius];
	\filldraw (tl) circle[radius=2pt];
        \filldraw (bl) circle[radius=2pt];
	\filldraw (r) circle[radius=2pt];
	\draw (tl) -- (etl);
	\draw (bl) -- (ebl);
	\draw (r) -- (etr);
 \draw (r) -- (ebr);
        \draw (tl) to[out=315, in=45] (bl);
  \end{tikzpicture}
\,, \qquad
    b_{2} = b_{2}^{(0)} + \varepsilon \, b_{2}^{(1)} \, ,
  \end{split}
\end{equation}
with coefficients given in Ref.~\cite{Benedetti:2020rrq}
\begin{equation}
\begin{split}
 b_{2}^{(0)}  &= 2\psi(d/4)-\psi(d/2)-\psi(1) \,, \\
 b_{2}^{(1)}  &=\frac{1}{4}\left\{2\psi(d/4)-\psi(d/2)-\psi(1)\right\}\left\{3\psi(1)-5\psi(d/2)+2\psi(d/4) \right\}
 \\&\qquad +\frac{3}{4}\psi_1(1)+\psi_1(d/4)-\frac{7}{4}\psi_1(d/2)-J_0(d/4) \,,
\end{split}
\end{equation}
where $J_0(d/4)$ is a complicate infinite summation of gamma and polygamma functions, see Ref.~\cite[Eq.\ (2.4)]{Benedetti:2020rrq}. For completeness we give the values for $d=2,3$ needed in this paper: $J_0(2/4)\simeq 0.374868$ and $J_0(3/4)\simeq 0.608524$.
The contribution to the metric is parametrized as
\begin{equation}\label{eq:metric2}
 \begin{split}
G^{(2)}_{IJ} &=
 m_{2,a}\,{\cal S}_{18}\,\left[
    \begin{tikzpicture}[scale=0.5,baseline=(vert_cent.base)]
	\node (vert_cent) at (0,0) {$\phantom{\cdot}$};
	\node (center) at (0,0) {};
	\def\vlength{0.5cm};
        \def\hlength{1cm}
        \node[inner sep=0pt] (l1) at (-1*\hlength,1.5*\vlength) {};
        \node[inner sep=0pt] (l2) at (-1*\hlength,0.5*\vlength) {};
	\node[inner sep=0pt] (l3) at (-1*\hlength,-0.5*\vlength) {};
 \node[inner sep=0pt] (l4) at (-1*\hlength,-1.5*\vlength) {};
	\node[inner sep=0pt] (r1) at (1*\hlength,1.5*\vlength) {};
	\node[inner sep=0pt] (r2) at (1*\hlength,0.5*\vlength){};
        \node[inner sep=0pt] (r3) at (1*\hlength,-0.5*\vlength) {};
        \node[inner sep=0pt] (r4) at (1*\hlength,-1.5*\vlength) {};
        \node[inner sep=0pt] (i1) at (0*\hlength,1*\vlength) {};
        \node[inner sep=0pt] (i2) at (0*\hlength,-1*\vlength) {};
	\filldraw (i1) circle[radius=2pt];
        \filldraw (i2) circle[radius=2pt];
        \draw (l1) -- (i1);
        \draw (l2) -- (i1);
        \draw (l3) -- (i2);
        \draw (l4) -- (i2);
         \draw (r1) -- (i1);
        \draw (r2) -- (i1);
        \draw (r3) -- (i2);
        \draw (r4) -- (i2);
    \end{tikzpicture}
    \right]
 +m_{2,b}\,{\cal S}_{16}\,\left[
   \begin{tikzpicture}[scale=0.5,baseline=(vert_cent.base)]
	\node (vert_cent) at (0,0) {$\phantom{\cdot}$};
	\node (center) at (0,0) {};
	\def\vlength{0.5cm};
        \def\hlength{1cm}
        \node[inner sep=0pt] (l1) at (-1*\hlength,1.5*\vlength) {};
        \node[inner sep=0pt] (l2) at (-1*\hlength,0.75*\vlength) {};
	\node[inner sep=0pt] (l3) at (-1*\hlength,0*\vlength) {};
 \node[inner sep=0pt] (l4) at (-1*\hlength,-0.75*\vlength) {};
	\node[inner sep=0pt] (r1) at (1*\hlength,0.75*\vlength) {};
	\node[inner sep=0pt] (r2) at (1*\hlength,0*\vlength){};
        \node[inner sep=0pt] (r3) at (1*\hlength,-0.75*\vlength) {};
        \node[inner sep=0pt] (r4) at (1*\hlength,-1.5*\vlength) {};
        \node[inner sep=0pt] (i1) at (0*\hlength,0.75*\vlength) {};
        \node[inner sep=0pt] (i2) at (0*\hlength,-0.75*\vlength) {};
	\filldraw (i1) circle[radius=2pt];
        \filldraw (i2) circle[radius=2pt];
        \draw (l1) -- (i1);
        \draw (l2) -- (i1);
        \draw (l3) -- (i1);
        \draw (l4) -- (i2);
         \draw (r1) -- (i1);
        \draw (r2) -- (i2);
        \draw (r3) -- (i2);
        \draw (r4) -- (i2);
    \end{tikzpicture}
    \right]
    +m_{2,c}\,{\cal S}_{144}\,\left[
       \begin{tikzpicture}[scale=0.5,baseline=(vert_cent.base)]
	\node (vert_cent) at (0,0) {$\phantom{\cdot}$};
	\node (center) at (0,0) {};
	\def\vlength{0.5cm};
        \def\hlength{1cm}
        \node[inner sep=0pt] (l1) at (-1*\hlength,0.75*\vlength) {};
        \node[inner sep=0pt] (l2) at (-1*\hlength,-0.25*\vlength) {};
	\node[inner sep=0pt] (l3) at (-1*\hlength,-1.25*\vlength) {};
 \node[inner sep=0pt] (l4) at (-1*\hlength,-1.5*\vlength) {};
	\node[inner sep=0pt] (r1) at (1*\hlength,1.25*\vlength) {};
	\node[inner sep=0pt] (r2) at (1*\hlength,0.25*\vlength){};
        \node[inner sep=0pt] (r3) at (1*\hlength,-0.75*\vlength) {};
        \node[inner sep=0pt] (r4) at (1*\hlength,-1.5*\vlength) {};
        \node[inner sep=0pt] (i1) at (0*\hlength,0.75*\vlength) {};
        \node[inner sep=0pt] (i2) at (0*\hlength,-0.75*\vlength) {};
	\filldraw (i1) circle[radius=2pt];
        \filldraw (i2) circle[radius=2pt];
        \draw (l1) -- (i1);
        \draw (l2) -- (i2);
        \draw (l3) -- (i2);
        \draw (l4) -- (r4);
         \draw (r1) -- (i1);
        \draw (r2) -- (i1);
        \draw (r3) -- (i2);
        \draw (i1) -- (i2);
    \end{tikzpicture}
    \right]
    \\&
    +m_{2,d}\,{\cal S}_{144}\,\left[
    \begin{tikzpicture}[scale=0.5,baseline=(vert_cent.base)]
	\node (vert_cent) at (0,0) {$\phantom{\cdot}$};
	\node (center) at (0,0) {};
	\def\vlength{0.5cm};
        \def\hlength{1cm}
        \node[inner sep=0pt] (l1) at (-1*\hlength,1.25*\vlength) {};
        \node[inner sep=0pt] (l2) at (-1*\hlength,-0.75*\vlength) {};
	\node[inner sep=0pt] (l3) at (-1*\hlength,-1*\vlength) {};
 \node[inner sep=0pt] (l4) at (-1*\hlength,-1.25*\vlength) {};
  \node[inner sep=0pt] (l5) at (-1*\hlength,-1.75*\vlength) {};
	\node[inner sep=0pt] (r1) at (1*\hlength,1.25*\vlength) {};
	\node[inner sep=0pt] (r2) at (1*\hlength,0.25*\vlength){};
        \node[inner sep=0pt] (r3) at (1*\hlength,-1*\vlength) {};
        \node[inner sep=0pt] (r4) at (1*\hlength,-1.25*\vlength) {};
        \node[inner sep=0pt] (r5) at (1*\hlength,-1.75*\vlength) {};
        \node[inner sep=0pt] (i1) at (0*\hlength,0.75*\vlength) {};
        \node[inner sep=0pt] (i2) at (0*\hlength,-0.5*\vlength) {};
	\filldraw (i1) circle[radius=2pt];
        \filldraw (i2) circle[radius=2pt];
        \draw (l1) -- (i1);
        \draw (l2) -- (i2);
        \draw (l4) -- (r4);
        \draw (l5) -- (r5);
         \draw (r1) -- (i1);
        \draw (r3) -- (i2);
        \draw (i1) to[out=-30, in=30] (i2);
        \draw (i1) to[out=210, in=150] (i2);
    \end{tikzpicture}
    \right]
    +m_{2,e}\,{\cal S}_{16}\,\left[
       \begin{tikzpicture}[scale=0.5,baseline=(vert_cent.base)]
	\node (vert_cent) at (0,0) {$\phantom{\cdot}$};
	\node (center) at (0,0) {};
	\def\vlength{0.5cm};
        \def\hlength{1cm}
        \node[inner sep=0pt] (l1) at (-1*\hlength,1.25*\vlength) {};
        \node[inner sep=0pt] (l2) at (-1*\hlength,0.5*\vlength) {};
	\node[inner sep=0pt] (l3) at (-1*\hlength,-0.25*\vlength) {};
 \node[inner sep=0pt] (l4) at (-1*\hlength,-0.75*\vlength) {};
	\node[inner sep=0pt] (r1) at (1*\hlength,1.25*\vlength) {};
	\node[inner sep=0pt] (r2) at (1*\hlength,0.5*\vlength){};
        \node[inner sep=0pt] (r3) at (1*\hlength,-0.25*\vlength) {};
        \node[inner sep=0pt] (r4) at (1*\hlength,-0.75*\vlength) {};
        \node[inner sep=0pt] (i1) at (-0.25*\hlength,0.5*\vlength) {};
        \node[inner sep=0pt] (i2) at (0.25*\hlength,0.5*\vlength) {};
	\filldraw (i1) circle[radius=2pt];
        \filldraw (i2) circle[radius=2pt];
        \draw (l1) -- (i1);
        \draw (l2) -- (i1);
        \draw (l3) -- (i1);
        \draw (l4) -- (r4);
         \draw (r1) -- (i2);
        \draw (r2) -- (i2);
        \draw (r3) -- (i2);
        \draw (i1) -- (i2);
    \end{tikzpicture}
    \right]
    +m_{2,f}\,{\cal S}_{72}\,\left[
     \begin{tikzpicture}[scale=0.5,baseline=(vert_cent.base)]
	\node (vert_cent) at (0,0) {$\phantom{\cdot}$};
	\node (center) at (0,0) {};
	\def\vlength{0.5cm};
        \def\hlength{1cm}
        \node[inner sep=0pt] (l1) at (-1*\hlength,1.5*\vlength) {};
        \node[inner sep=0pt] (l2) at (-1*\hlength,0*\vlength) {};
	\node[inner sep=0pt] (l3) at (-1*\hlength,-0.5*\vlength) {};
 \node[inner sep=0pt] (l4) at (-1*\hlength,-1.25*\vlength) {};
	\node[inner sep=0pt] (r1) at (1*\hlength,1.5*\vlength) {};
	\node[inner sep=0pt] (r2) at (1*\hlength,0*\vlength){};
        \node[inner sep=0pt] (r3) at (1*\hlength,-0.5*\vlength) {};
        \node[inner sep=0pt] (r4) at (1*\hlength,-1.25*\vlength) {};
        \node[inner sep=0pt] (i1) at (-0.5*\hlength,0.75*\vlength) {};
        \node[inner sep=0pt] (i2) at (0.5*\hlength,0.75*\vlength) {};
	\filldraw (i1) circle[radius=2pt];
        \filldraw (i2) circle[radius=2pt];
        \draw (l1) -- (i1);
        \draw (l2) -- (i1);
        \draw (l3) -- (r3);
        \draw (l4) -- (r4);
         \draw (r1) -- (i2);
        \draw (r2) -- (i2);
        \draw (i1) to[out=45, in=135]  (i2);
        \draw (i1) to[out=-45, in=-135]  (i2);
    \end{tikzpicture}
    \right]
\\&
    +m_{2,g}\,{\cal S}_{96}\,\left[
       \begin{tikzpicture}[scale=0.5,baseline=(vert_cent.base)]
	\node (vert_cent) at (0,0) {$\phantom{\cdot}$};
	\node (center) at (0,0) {};
	\def\vlength{0.5cm};
        \def\hlength{1cm}
        \node[inner sep=0pt] (l1) at (-1*\hlength,1.25*\vlength) {};
        \node[inner sep=0pt] (l2) at (-1*\hlength,0.5*\vlength) {};
	\node[inner sep=0pt] (l3) at (-1*\hlength,-0.25*\vlength) {};
 \node[inner sep=0pt] (l4) at (-1*\hlength,-1*\vlength) {};
	\node[inner sep=0pt] (r1) at (1*\hlength,1.25*\vlength) {};
	\node[inner sep=0pt] (r2) at (1*\hlength,0.5*\vlength){};
        \node[inner sep=0pt] (r3) at (1*\hlength,-0.25*\vlength) {};
        \node[inner sep=0pt] (r4) at (1*\hlength,-1*\vlength) {};
        \node[inner sep=0pt] (i1) at (-0.5*\hlength,0.5*\vlength) {};
        \node[inner sep=0pt] (i2) at (0.5*\hlength,0.5*\vlength) {};
	\filldraw (i1) circle[radius=2pt];
        \filldraw (i2) circle[radius=2pt];
        \draw (l1) -- (r1);
        \draw (l2) -- (i1);
        \draw (i2) -- (r2);
        \draw (l3) -- (r3);
        \draw (l4) -- (r4);
        \draw (i1) -- (i2);
         \draw (i1) to[out=45, in=135]  (i2);
        \draw (i1) to[out=-45, in=-135]  (i2);
    \end{tikzpicture}
    \right] \, ,
 \end{split}
\end{equation}
where the $m$s should be determined at the leading order $\varepsilon^0$. The contribution to the $A$-function is parametrized as
\begin{equation}\label{eq:afun2}
 \begin{split}
 A^{(2)} &= 
 a_{2,a}\,
  \begin{tikzpicture}[scale=0.5,baseline=(vert_cent.base)]
	\node (vert_cent) at (0,0) {$\phantom{\cdot}$};
	\node (center) at (0,0) {};
	\def\radius{1cm};
	\node[inner sep=0pt] (tr) at (45:\radius) {};
	\node[inner sep=0pt] (tl) at (135:\radius) {};
	\node[inner sep=0pt] (bl) at (225:\radius) {};
        \node[inner sep=0pt] (br) at (315:\radius) {};
	\draw (center) circle[radius=\radius];
	\filldraw (tr) circle[radius=2pt];
        \filldraw (tl) circle[radius=2pt];
        \filldraw (bl) circle[radius=2pt];
        \filldraw (br) circle[radius=2pt];
	\draw (tr) -- (bl);
	\draw (tl) -- (br);
 \draw (tl) to[out=-45, in=225] (tr);
 \draw (bl) to[out=45, in=135] (br);
	\end{tikzpicture}
+a_{2,b}\,
  \begin{tikzpicture}[scale=0.5,baseline=(vert_cent.base)]
	\node (vert_cent) at (0,0) {$\phantom{\cdot}$};
	\node (center) at (0,0) {};
	\def\radius{1cm};
	\node[inner sep=0pt] (tr) at (45:\radius) {};
	\node[inner sep=0pt] (tl) at (135:\radius) {};
	\node[inner sep=0pt] (bl) at (225:\radius) {};
        \node[inner sep=0pt] (br) at (315:\radius) {};
	\draw (center) circle[radius=\radius];
	\filldraw (tr) circle[radius=2pt];
        \filldraw (tl) circle[radius=2pt];
        \filldraw (bl) circle[radius=2pt];
        \filldraw (br) circle[radius=2pt];
	\draw (tr) to[out=225, in=135] (br);
	\draw (tl) to[out=-45, in=45] (bl);
  \draw (tl) to[out=-45, in=225] (tr);
 \draw (bl) to[out=45, in=135] (br);
    \end{tikzpicture}
    +a_{2,c}\,
  \begin{tikzpicture}[scale=0.5,baseline=(vert_cent.base)]
	\node (vert_cent) at (0,0) {$\phantom{\cdot}$};
	\node (center) at (0,0) {};
	\def\radius{1cm};
	\node[inner sep=0pt] (tr) at (45:\radius) {};
	\node[inner sep=0pt] (tl) at (135:\radius) {};
	\node[inner sep=0pt] (bl) at (225:\radius) {};
        \node[inner sep=0pt] (br) at (315:\radius) {};
	\draw (center) circle[radius=\radius];
	\filldraw (tr) circle[radius=2pt];
        \filldraw (tl) circle[radius=2pt];
        \filldraw (bl) circle[radius=2pt];
        \filldraw (br) circle[radius=2pt];
  \draw (tl) to[out=-45, in=225] (tr);
  \draw (tl) -- (tr);
 \draw (bl) to[out=45, in=135] (br);
 \draw (bl) -- (br);
    \end{tikzpicture}
 \end{split}
\,, \qquad
    a_{2,i} = a_{2,i}^{(0)} + \varepsilon \, a_{2,i}^{(1)} \, .
\end{equation}
The solution to the gradient equation introduces one new free parameter, which we choose as $\alpha_4=m_{2,a}^{(0)}$, resulting in
\begin{equation}\label{eq:consistency-2loop}
\begin{split}
    & a_{2,a}^{(0)} = \frac{\alpha_2 b_1^{(0)}}{2}+\frac{3b_{2}^{(0)}}{2}
    \, , \qquad
    a_{2,b}^{(0)} = \frac{\alpha_2 b_1^{(0)}}{4}
    \, , \qquad
    a_{2,c}^{(0)} = 0  \, .
\end{split}
\end{equation}
Some $A$ and $G$ coefficients are partly undetermined:
the subleading coefficients $a^{(1)}_{2,i}$ for $i=a,b,c$ in Eq.~\eqref{eq:afun2}
and the leading coefficients $m^{(0)}_{2,i}$ for $i=a,b,\cdots,g$ in Eq.~\eqref{eq:metric2} can be found in the next subsection, as the solution at that order depends on the coefficients of the beta functions at the next loop order.

\subsection{$3$-loop consistency}\label{sect:3loop}

The new terms in the beta function are
\begin{equation}\label{eq:beta3}
 \begin{split}
  \beta^{(3)}_{ijkl} &=
  b_{3,a}\,{\cal S}_1\,
  \begin{tikzpicture}[scale=0.5,baseline=(vert_cent.base)]
	\node (vert_cent) at (0,0) {$\phantom{\cdot}$};
	\node (center) at (0,0) {};
	\def\radius{1cm};
    \node[inner sep=0pt] (it) at (150:1.5\radius) {};
    \node[inner sep=0pt] (ib) at (210:1.5\radius) {};
    \node[inner sep=0pt] (ot) at (30:1.5\radius) {};
    \node[inner sep=0pt] (ob) at (330:1.5\radius) {};
	\node[inner sep=0pt] (tr) at (45:\radius) {};
	\node[inner sep=0pt] (tl) at (135:\radius) {};
    \node[inner sep=0pt] (br) at (315:\radius) {};
    \node[inner sep=0pt] (bl) at (225:\radius) {};
	\draw (center) circle[radius=\radius];
	    \filldraw (tr) circle[radius=2pt];
        \filldraw (tl) circle[radius=2pt];
        \filldraw (bl) circle[radius=2pt];
        \filldraw (br) circle[radius=2pt];
    \draw (it) -- (tl);
    \draw (ib) -- (bl);
    \draw (ot) -- (tr);
    \draw (ob) -- (br);
	\draw (tl) -- (br);
    \draw (tr) -- (bl);
	\end{tikzpicture}
 + b_{3,b}\,{\cal S}_3\,
  \begin{tikzpicture}[scale=0.5,baseline=(vert_cent.base)]
	\node (vert_cent) at (0,0) {$\phantom{\cdot}$};
	\node (center) at (0,0) {};
	\def\radius{1cm};
    \node[inner sep=0pt] (it) at (150:1.5\radius) {};
    \node[inner sep=0pt] (ib) at (220:1.5\radius) {};
    \node[inner sep=0pt] (ot) at (30:1.5\radius) {};
    \node[inner sep=0pt] (ob) at (330:1.5\radius) {};
	\node[inner sep=0pt] (r) at (0:\radius) {};
    \node[inner sep=0pt] (l) at (180:\radius) {};
    \node[inner sep=0pt] (tr) at (45:0.8\radius) {};
    \node[inner sep=0pt] (tl) at (135:0.8\radius) {};
	\filldraw (r) circle[radius=2pt];
    \filldraw (l) circle[radius=2pt];
	\filldraw (tr) circle[radius=2pt];
    \filldraw (tl) circle[radius=2pt];
    \draw (it) -- (l);
    \draw (ib) -- (l);
    \draw (ot) -- (r);
    \draw (ob) -- (r);
	\draw (l) to[out=270, in=270] (r);
	\draw (l) to[out=90, in=180] (tl);
	\draw (tl) -- (tr);
    \draw (tl)  to[out=45, in=135] (tr);
    \draw (tl) to[out=315, in=225] (tr);
    \draw (tr) to[out=0, in=90] (r);
  \end{tikzpicture}
  +b_{3,c}\,{\cal S}_3\,
    \begin{tikzpicture}[scale=0.5,baseline=(vert_cent.base)]
	\node (vert_cent) at (0,0) {$\phantom{\cdot}$};
	\node (center) at (0,0) {};
	\def\radius{1cm};
    \node[inner sep=0pt] (it) at (150:1.5\radius) {};
    \node[inner sep=0pt] (ib) at (220:1.5\radius) {};
    \node[inner sep=0pt] (ot) at (30:1.5\radius) {};
    \node[inner sep=0pt] (ob) at (330:1.5\radius) {};
	\node[inner sep=0pt] (r) at (0:\radius) {};
    \node[inner sep=0pt] (l) at (180:\radius) {};
    \node[inner sep=0pt] (t) at (90:0.6\radius) {};
    \node[inner sep=0pt] (b) at (270:0.6\radius) {};
	\filldraw (r) circle[radius=2pt];
    \filldraw (l) circle[radius=2pt];
	\filldraw (t) circle[radius=2pt];
    \filldraw (b) circle[radius=2pt];
    \draw (it) -- (l);
    \draw (ib) -- (l);
    \draw (ot) -- (r);
    \draw (ob) -- (r);
	\draw (l) to[out=270, in=270] (r);
	\draw (l) to[out=90, in=90] (r);
    \draw (t) to[out=300, in=60] (b);
    \draw (t) to[out=240, in=120] (b);
  \end{tikzpicture}
     \\
     &
     +b_{3,d}\,{\cal S}_6\,
    \begin{tikzpicture}[scale=0.5,baseline=(vert_cent.base)]
	\node (vert_cent) at (0,0) {$\phantom{\cdot}$};
	\node (center) at (0,0) {};
	\def\radius{1cm};
    \node[inner sep=0pt] (it) at (150:1.5\radius) {};
    \node[inner sep=0pt] (ib) at (210:1.5\radius) {};
    \node[inner sep=0pt] (ot) at (30:1.5\radius) {};
    \node[inner sep=0pt] (ob) at (330:1.5\radius) {};
	\node[inner sep=0pt] (r) at (0:\radius) {};
    \node[inner sep=0pt] (l) at (180:\radius) {};
    \node[inner sep=0pt] (tr) at (60:0.7\radius) {};
    \node[inner sep=0pt] (br) at (300:0.7\radius) {};
	\filldraw (r) circle[radius=2pt];
    \filldraw (l) circle[radius=2pt];
	\filldraw (tr) circle[radius=2pt];
    \filldraw (br) circle[radius=2pt];
    \draw (it) -- (l);
    \draw (ib) -- (l);
    \draw (ot) -- (tr);
    \draw (ob) -- (br);
	\draw (tr) to[out=270, in=160] (r);
	\draw (r) to[out=200, in=90] (br);
    \draw (l) to[out=270, in=270] (r);
    \draw (l) to[out=90, in=90] (r);
  \end{tikzpicture} 
     +b_{3,e}\,{\cal S}_6\,
  \begin{tikzpicture}[scale=0.5,baseline=(vert_cent.base)]
	\node (vert_cent) at (0,0) {$\phantom{\cdot}$};
	\node (center) at (0,0) {};
	\def\radius{1cm};
    \node[inner sep=0pt] (it) at (150:1.5\radius) {};
    \node[inner sep=0pt] (ib) at (210:1.5\radius) {};
    \node[inner sep=0pt] (ot) at (30:1.5\radius) {};
    \node[inner sep=0pt] (ob) at (330:1.5\radius) {};
	\node[inner sep=0pt] (tr) at (45:\radius) {};
	\node[inner sep=0pt] (tl) at (135:\radius) {};
    \node[inner sep=0pt] (br) at (315:\radius) {};
    \node[inner sep=0pt] (bl) at (225:\radius) {};
	\draw (center) circle[radius=\radius];
	    \filldraw (tr) circle[radius=2pt];
        \filldraw (tl) circle[radius=2pt];
        \filldraw (bl) circle[radius=2pt];
        \filldraw (br) circle[radius=2pt];
    \draw (it) -- (tl);
    \draw (ib) -- (bl);
    \draw (ot) -- (tr);
    \draw (ob) -- (br);
	\draw (tl) to[out=315, in=45] (bl);
    \draw (tr) to[out=225, in=125] (br);
	\end{tikzpicture}
    +b_{3,f}\,{\cal S}_{12}\,
  \begin{tikzpicture}[scale=0.5,baseline=(vert_cent.base)]
	\node (vert_cent) at (0,0) {$\phantom{\cdot}$};
	\node (center) at (0,0) {};
	\def\radius{1cm};
    \node[inner sep=0pt] (it) at (150:1.5\radius) {};
    \node[inner sep=0pt] (ib) at (210:1.5\radius) {};
    \node[inner sep=0pt] (ot) at (30:1.5\radius) {};
    \node[inner sep=0pt] (ob) at (330:1.5\radius) {};
	\node[inner sep=0pt] (tr) at (45:\radius) {};
	\node[inner sep=0pt] (tl) at (135:\radius) {};
    \node[inner sep=0pt] (br) at (315:\radius) {};
    \node[inner sep=0pt] (l) at (180:\radius) {};
	\draw (center) circle[radius=\radius];
	    \filldraw (tr) circle[radius=2pt];
        \filldraw (tl) circle[radius=2pt];
        \filldraw (l) circle[radius=2pt];
        \filldraw (br) circle[radius=2pt];
    \draw (it) -- (l);
    \draw (ib) -- (l);
    \draw (ot) -- (tr);
    \draw (ob) -- (br);
	\draw (tl) to[out=330, in=210] (tr);
    \draw (tl) -- (br);
	\end{tikzpicture}
   \ ,
  \end{split}
\end{equation}
with coefficients determined in Ref.~\cite{Benedetti:2020rrq}
\begin{equation}
\begin{split}
    & b_{3,b}= \frac{\Gamma\left( -d/4 \right)\Gamma \left( d/2\right)^2}{3 \Gamma \left( 3d/4\right)}  
    \,, \\
    & b_{3,c}=\frac{1}{2}(2 \psi(d/4)-\psi(d/2)-\psi(1))^2+\frac{1}{2}\psi_1(1)+\psi_1(d/4)-\frac{3}{2}\psi_1(d/2)-J_0(d/4)
    \,, \\
    & b_{3,d}= -\psi_1(1) - \psi_1(d/4)+2 \psi_1(d/2)+J_0(d/4)
    \,, \\
    & b_{3,e}= -\psi_1(1) - \psi_1(d/4)+2 \psi_1(d/2)+J_0(d/4)
    \,, \\
    &  b_{3,f}= \frac{3}{2} (2 \psi(d/4)-\psi(d/2)-\psi(1))^2+\frac{1}{2}\psi_1(1)-\frac{1}{2}\psi_1(d/2) \, .
\end{split}
\end{equation}
The $b_{3,a}$ coefficient is a complicate double sum given in Refs.~\cite{Benedetti_2025,Fraser-Taliente:2026iuj}, of which we only give the numerical values in $d=2,3$ for simplicity
\begin{equation}
    b_{3,a}\simeq\ \left \{ \begin{array}{ll}
        76.62828703(7) \quad & \text{in $d=2$} \,, \\
        30.1026152(6)  \quad & \text{in $d=3$} \,.
    \end{array} \right.
\end{equation}

The leading terms of the metric $G^{(3)}_{IJ}$ affect only the next order, so they are not necessary here. The $A$-function now includes
\begin{equation}\label{eq:afun3}
 \begin{split}
 A^{(3)} &= 
 a_{3,a}\,
  \begin{tikzpicture}[scale=0.5,baseline=(vert_cent.base)]
	\node (vert_cent) at (0,0) {$\phantom{\cdot}$};
	\node (center) at (0,0) {};
	\def\radius{1cm};
	\node[inner sep=0pt] (t) at (90:\radius) {};
	\node[inner sep=0pt] (tr) at (18:\radius) {};
	\node[inner sep=0pt] (tl) at (162:\radius) {};
    \node[inner sep=0pt] (br) at (306:\radius) {};
    \node[inner sep=0pt] (bl) at (234:\radius) {};
	\draw (center) circle[radius=\radius];
        \filldraw (t) circle[radius=2pt];
	    \filldraw (tr) circle[radius=2pt];
        \filldraw (tl) circle[radius=2pt];
        \filldraw (bl) circle[radius=2pt];
        \filldraw (br) circle[radius=2pt];
	\draw (t) -- (bl);
    \draw (t) -- (br);
	\draw (tl) -- (tr);
    \draw (tr) -- (bl);
    \draw (tl) -- (br);
	\end{tikzpicture}
+a_{3,b}\,
  \begin{tikzpicture}[scale=0.5,baseline=(vert_cent.base)]
	\node (vert_cent) at (0,0) {$\phantom{\cdot}$};
	\node (center) at (0,0) {};
	\def\radius{1cm};
	\node[inner sep=0pt] (tr) at (45:\radius) {};
	\node[inner sep=0pt] (tl) at (135:\radius) {};
	\node[inner sep=0pt] (bl) at (225:\radius) {};
    \node[inner sep=0pt] (br) at (315:\radius) {};
	\draw (center) circle[radius=\radius];
        \filldraw (center) circle[radius=2pt];
	    \filldraw (tr) circle[radius=2pt];
        \filldraw (tl) circle[radius=2pt];
        \filldraw (bl) circle[radius=2pt];
        \filldraw (br) circle[radius=2pt];
	\draw (tl) -- (br);
    \draw (tr) -- (bl);
    \draw (tr) to[out=240, in=120] (br);
    \draw (tl) to[out=300, in=60] (bl);
	\end{tikzpicture}
    +a_{3,c}\,
   \begin{tikzpicture}[scale=0.5,baseline=(vert_cent.base)]
	\node (vert_cent) at (0,0) {$\phantom{\cdot}$};
	\node (center) at (0,0) {};
	\def\radius{1cm};
	\node[inner sep=0pt] (t) at (90:\radius) {};
	\node[inner sep=0pt] (tr) at (18:\radius) {};
	\node[inner sep=0pt] (tl) at (162:\radius) {};
    \node[inner sep=0pt] (br) at (306:\radius) {};
    \node[inner sep=0pt] (bl) at (234:\radius) {};
	\draw (center) circle[radius=\radius];
        \filldraw (t) circle[radius=2pt];
	    \filldraw (tr) circle[radius=2pt];
        \filldraw (tl) circle[radius=2pt];
        \filldraw (bl) circle[radius=2pt];
        \filldraw (br) circle[radius=2pt];
	\draw (t) -- (bl);
    \draw (tl) -- (br);
    \draw (tl) to[out=310, in=100] (bl);
    \draw (t) to[out=305, in=165] (tr);
    \draw (tr) to[out=215, in=105] (br);
	\end{tikzpicture}
    +a_{3,d}\,
   \begin{tikzpicture}[scale=0.5,baseline=(vert_cent.base)]
	\node (vert_cent) at (0,0) {$\phantom{\cdot}$};
	\node (center) at (0,0) {};
	\def\radius{1cm};
	\node[inner sep=0pt] (t) at (90:\radius) {};
	\node[inner sep=0pt] (tr) at (18:\radius) {};
	\node[inner sep=0pt] (tl) at (162:\radius) {};
    \node[inner sep=0pt] (br) at (306:\radius) {};
    \node[inner sep=0pt] (bl) at (234:\radius) {};
	\draw (center) circle[radius=\radius];
        \filldraw (t) circle[radius=2pt];
	    \filldraw (tr) circle[radius=2pt];
        \filldraw (tl) circle[radius=2pt];
        \filldraw (bl) circle[radius=2pt];
        \filldraw (br) circle[radius=2pt];
	\draw (t) to[out=240, in=20] (tl);
    \draw (bl) to[out=30, in=145] (br);
    \draw (tl) to[out=320, in=90] (bl);
    \draw (t) to[out=305, in=165] (tr);
    \draw (tr) to[out=215, in=105] (br);
	\end{tikzpicture}
    +a_{3,e}\,
   \begin{tikzpicture}[scale=0.5,baseline=(vert_cent.base)]
	\node (vert_cent) at (0,0) {$\phantom{\cdot}$};
	\node (center) at (0,0) {};
	\def\radius{1cm};
	\node[inner sep=0pt] (t) at (90:\radius) {};
	\node[inner sep=0pt] (tr) at (18:\radius) {};
	\node[inner sep=0pt] (tl) at (162:\radius) {};
    \node[inner sep=0pt] (br) at (306:\radius) {};
    \node[inner sep=0pt] (bl) at (234:\radius) {};
	\draw (center) circle[radius=\radius];
        \filldraw (t) circle[radius=2pt];
	    \filldraw (tr) circle[radius=2pt];
        \filldraw (tl) circle[radius=2pt];
        \filldraw (bl) circle[radius=2pt];
        \filldraw (br) circle[radius=2pt];
	\draw (tl) -- (bl);
    \draw (t) -- (br);
    \draw (tl) to[out=330, in=70] (bl);
    \draw (t) to[out=305, in=165] (tr);
    \draw (tr) to[out=215, in=105] (br);
	\end{tikzpicture}
 \end{split}
\, ,
\end{equation}
and we can only determine consistently the leading contributions to its coefficients, i.e.\ $a_{3,i} = a_{3,i}^{(0)}$ for $i=a,b,\cdots,e$.

For the solution we introduce $\alpha_5=m_{2,b}^{(0)}$, which is our last free parameter of the consistency conditions. The still undetermined parameters from the previous subsection are
\begin{equation}\label{eq:consistency-3loop}
\begin{split}
 a_{2,a}^{(1)} &=
  -\frac{5\alpha_4}{4}
  +\frac{\alpha_3 b_1^{(0)}}{2}
  -\frac{\alpha_2 (-2 b_1^{(0)} b_1^{(1)}+3 b_{2}^{(0)})}{4 b_1^{(0)}}
  \\& 
  +\frac{3 (4 b_1^{(0)} b_{2}^{(1)}-5 b_{3,c}^{(0)}+5 b_{3,d}^{(0)}-b_{3,f}^{(0)} +2 b_{3,e}^{(0)})}{8 b_1^{(0)}}+\frac{3 \alpha_0 b_{2}^{(0)}}{2} \, ,
  \\
  a_{2,b}^{(1)} &= -\frac{5 \alpha_4}{8}+\frac{\alpha_3 b_1^{(0)}}{4}+\frac{\alpha_2  b_1^{(1)}}{4} +\frac{3 (-3 b_{3,c}^{(0)}-b_{3,d}^{(0)}+b_{3,f}^{(0)}-2 b_{3,e}^{(0)})}{16 b_1^{(0)}} \, ,
  \\
  a_{2,c}^{(1)} &= -\frac{b_{3,b}^{(0)}}{b_1^{(0)}} \, ,
\end{split}
\end{equation}
and
\begin{equation}\label{eq:consistency-3loop-2}
\begin{split}
 &m_{2,c}^{(0)} = 4 \alpha_4+\frac{2 \alpha_2 b_{2}^{(0)}}{b_1^{(0)}}+\frac{6 (b_{3,c}^{(0)}-b_{3,d}^{(0)})}{b_1^{(0)}} \ ,  \\
        & m_{2,d}^{(0)} = \alpha_4+\frac{\alpha_2 b_{2}^{(0)}}{b_1^{(0)}}+\frac{3 (b_{3,c}^{(0)}-b_{3,d}^{(0)}+b_{3,f}^{(0)}-2 b_{3,e}^{(0)})}{2 b_1^{(0)}}  \ , \\
        & m_{2,e}^{(0)} = \frac{2 b_{3,b}^{(0)}}{b_1^{(0)}}-\alpha_5  \ , \\
        & m_{2,f}^{(0)} = \frac{3 \alpha_4}{2}+\frac{3 (3 b_{3,c}^{(0)}+ b_{3,d}^{(0)}-b_{3,f}^{(0)}+2 b_{3,e}^{(0)})}{4 b_1^{(0)}} \ ,  \\
        & m_{2,g}^{(0)} = \frac{2 b_{3,b}^{(0)}}{b_1^{(0)}} \, .
\end{split}
\end{equation}
The new coefficients of $A$ are
\begin{equation}\label{eq:consistency-3loop-3}
\begin{split}
    & a_{3,a}^{(0)} = \frac{b_{3,a}^{(0)}}{5} \ , \\
        & a_{3,b}^{(0)} = 2 \alpha_4 b_1^{(0)}+2 \alpha_2 b_{2}^{(0)}+3 (b_{3,c}^{(0)}-b_{3,d}^{(0)}+b_{3,f}^{(0)}) \ ,  \\
        & a_{3,c}^{(0)} = 2 \alpha_4 b_1^{(0)}+\alpha_2 b_{2}^{(0)}+3 b_{3,c}^{(0)}  \ , \\
        & a_{3,d}^{(0)} = \frac{\alpha_4 b_1^{(0)}}{2}+\frac{3}{20} (3 b_{3,c}^{(0)}+b_{3,d}^{(0)}-b_{3,f}^{(0)}+2 b_{3,e}^{(0)}) \ ,  \\
        & a_{3,e}^{(0)} = 3 b_{3,b}^{(0)}  \, .
\end{split}
\end{equation}
This concludes the proof by inspection that the beta functions are integrable in the $\varepsilon$-expansion up to order $L'=3$.

\section{The $\tilde{F}$-function}\label{sect:F-function}

In order to match the RG quantities with their analogues in the CFT at the fixed point, we first need to compute the $\Tilde{F}$-function \cite{Giombi:2014xxa,Giombi:2015haa,Fei:2015oha} in this section, while the actual matching of $\Tilde{F}$ with $A$ is performed in Sect.~\ref{sect:matching}.

We warn the Reader that in this section and in the following ones we do not rescale the coupling, i.e.\ $\lambda_I \to \alpha \, \lambda_I$ for $\alpha=1$ in the action \eqref{eq:action}, differently than what has been done in Eq.~\eqref{eq:rescaling-benedetti} of Sect.~\ref{sect:consistency}. We have chosen to do so in order to simplify the comparison between Refs.~\cite{Benedetti:2020rrq, Giombi:2024zrt} that use different rescalings. Physical quantities in the $\varepsilon$-expansion are not affected by the rescaling, but fixed point values are and this will be important later.

Following Cardy \cite{Cardy:1988cwa}, the $\Tilde{F}$-function is a thermodynamical quantity computed by putting the field theory on the $d$-dimensional sphere $S^d$ of radius $R$. It is defined as
\begin{equation}\label{eq:F-tilde}
    \tilde{F}=\sin\left( \frac{\pi d}{2} \right) \log \left( Z_{S^d} \right) \, ,
\end{equation}
which interpolates with the conformal anomaly in even dimensions and the monotonic $F$-functions in odd dimension \cite{Giombi:2014xxa}. $F\equiv \log\left( Z_{S^d} \right)$ is the actual free-energy on the $d$-dimensional sphere, but we simply refer to $\tilde{F}=\sin\left( \frac{\pi d}{2} \right) F$ as the sphere free-energy, even if it is actually the free-energy up to a constant at fixed dimension.

Our aim is to compute $\Tilde{F}$ perturbatively for the long-range $\phi^4$ model \eqref{eq:action} in the $\varepsilon$-expansion on the $d$-dimensional sphere in the physically interesting dimensions $d=2$ and $d=3$ (see Fig.~\ref{fig:LRI}).
Our results can be checked against the existing literature, recently updated in Ref.~\cite[arXiv version]{Giombi:2024zrt}. Unlike the previous section, it is not straightforward to perform all computations in general $d$, which is why we restrict to $d=2,3$.

The first step is to couple the action \eqref{eq:action} to the sphere's metric
in a Weyl-covariant manner. This step has the additional complication caused by the fact that we should be working with a fractional Weyl-covariant Laplacian, but it can be done consistently
as elaborated in Ref.~\cite{Fraser-Taliente:2026iuj} for $\frac{d-2}{2} < \Delta_\phi < \frac{d}{2}$
which implies $d<4-\varepsilon$ and includes our cases of interest ($d=2,3$ for $\varepsilon$ small).
Then the $\tilde{F}$ function can be written as the sum of the term $\tilde{F}_{\rm Free}$ relative to the conformally coupled (generalized) free fields and an additional term $\delta \tilde{F}$ given by the local interactions of action \eqref{eq:action}.
For this paper we are only interested in the calculation of the latter,
but, for completeness, we report here the expression for the free part as a function of $s$ in general $d$ given in Ref.~\cite{Giombi:2014xxa}:
\begin{equation}\label{eq:F-Free-Fields}
    \tilde{F}_{\text{Free}}=\frac{1}{\Gamma(1+d)}\int_0^{\frac{s}{2}} {\rm d}u \ u \ \sin(\pi u) \ \Gamma \left( \frac{d}{2}+u \right)  \Gamma \left( \frac{d}{2}-u \right) \, .
\end{equation}
For $s=2$ we can recover the SR version, of which we give the solution in the integer dimensions of interest,
$$  \tilde{F}_{\text{Free}}\rvert_{d=2}=\frac{\zeta(2)}{\pi}=\frac{\pi}{6} \, , \quad \tilde{F}_{\text{Free}}\rvert_{d=3}=\frac{1}{16}\left( \log(4)-\frac{3 \zeta(3)}{\pi^2} \right) \, , $$
where $\zeta$ is the usual Riemann Zeta function.

In this and the next section the diagrams are obtained from promoting the action \eqref{eq:action} Weyl-covariantly to $S^d$ and represent actual perturbative Feynman diagrams computed in coordinate space, rather than simply contractions of the \emph{bare} coupling tensor. The results will be finite when expressed in terms of the \emph{renormalized} coupling tensor because $\tilde{F}$ does not require further renormalization.

We use the following notation for the scalar propagator in real space,
\begin{equation}\label{eq:propagator}
  \begin{tikzpicture}[scale=0.5,baseline=(vert_cent.base)]
	\node (vert_cent) at (0,0) {$\phantom{\cdot}$};
	\node (center) at (0,0) {};
	\def\radius{1cm};
    \node[inner sep=0pt] (tl) at (180:\radius) {};
    \node[inner sep=0pt] (tl1) at (180:1.5\radius) {$x$};
	\node[inner sep=0pt] (tr) at (0:\radius) {};
    \node[inner sep=0pt] (tr1) at (0:1.5\radius) {$y$};
        \filldraw (tl) circle[radius=2pt];
        \filldraw (tr) circle[radius=2pt];
	\draw (tr) -- (tl);
	\end{tikzpicture}
 \, 
 =  \frac{C_{d,s}}{\sigma(x,y)^{d-s}} \,, \qquad C_{d,s}=\frac{\Gamma(\frac{d-s}{2})}{\pi^{\frac{d}{2}}2^s\Gamma(\frac{s}{2})} \ ,
\end{equation}
where $\sigma(x,y)$ is the chordal distance on $S^d$.\footnote{%
A warning to avoid potential confusions: in applications of QFT in curved space notation $\sigma(x,y)$ is generally reserved for the bi-scalar known as the de Witt world function, which is one half of the square of the geodesic distance \cite{DeWitt:1964mxt}. In this paper it is instead simply the geodesic distance.
}
Following the prescriptions in Refs.~\cite{Giombi:2014xxa,Giombi:2024zrt}, the interaction part of the free-energy in general dimensions takes the form 
\begin{equation}\label{eq:F-general}
    \delta \tilde{F}= -\frac{1}{2! \ 4^2} \ \begin{tikzpicture}[scale=0.5,baseline=(vert_cent.base)]
	\node (vert_cent) at (0,0) {$\phantom{\cdot}$};
	\node (center) at (0,0) {};
	\def\radius{1cm};
	\node[inner sep=0pt] (l) at (0:\radius) {};
	\node[inner sep=0pt] (r) at (180:\radius) {};
	\draw (center) circle[radius=\radius];
	\filldraw (r) circle[radius=2pt];
    \filldraw (l) circle[radius=2pt];
	\draw (r) to[out=45, in=135] (l);
 \draw (r) to[out=-45, in=-135] (l);
	\end{tikzpicture} + \ \frac{1}{3! \ 4^3} \begin{tikzpicture}[scale=0.5,baseline=(vert_cent.base)]
	\node (vert_cent) at (0,0) {$\phantom{\cdot}$};
	\node (center) at (0,0) {};
	\def\radius{1cm};
	\node[inner sep=0pt] (tr) at (0:\radius) {};
	\node[inner sep=0pt] (tl) at (120:\radius) {};
	\node[inner sep=0pt] (bl) at (240:\radius) {};
	\draw (center) circle[radius=\radius];
	\filldraw (tr) circle[radius=2pt];
        \filldraw (tl) circle[radius=2pt];
        \filldraw (bl) circle[radius=2pt];
	\draw (tr) to[out=180, in=-60] (tl);
	\draw (tl) to[out=-60, in=60] (bl);
 \draw (tr) to[out=180, in=60] (bl);
	\end{tikzpicture} -  \frac{1}{4! \ 4^4} \ \left(  
  \begin{tikzpicture}[scale=0.5,baseline=(vert_cent.base)]
	\node (vert_cent) at (0,0) {$\phantom{\cdot}$};
	\node (center) at (0,0) {};
	\def\radius{1cm};
	\node[inner sep=0pt] (tr) at (45:\radius) {};
	\node[inner sep=0pt] (tl) at (135:\radius) {};
	\node[inner sep=0pt] (bl) at (225:\radius) {};
        \node[inner sep=0pt] (br) at (315:\radius) {};
	\draw (center) circle[radius=\radius];
	\filldraw (tr) circle[radius=2pt];
        \filldraw (tl) circle[radius=2pt];
        \filldraw (bl) circle[radius=2pt];
        \filldraw (br) circle[radius=2pt];
	\draw (tr) to[out=225, in=135] (br);
	\draw (tl) to[out=-45, in=45] (bl);
  \draw (tl) to[out=-45, in=225] (tr);
 \draw (bl) to[out=45, in=135] (br);
    \end{tikzpicture}
    +
  \begin{tikzpicture}[scale=0.5,baseline=(vert_cent.base)]
	\node (vert_cent) at (0,0) {$\phantom{\cdot}$};
	\node (center) at (0,0) {};
	\def\radius{1cm};
	\node[inner sep=0pt] (tr) at (45:\radius) {};
	\node[inner sep=0pt] (tl) at (135:\radius) {};
	\node[inner sep=0pt] (bl) at (225:\radius) {};
        \node[inner sep=0pt] (br) at (315:\radius) {};
	\draw (center) circle[radius=\radius];
	\filldraw (tr) circle[radius=2pt];
        \filldraw (tl) circle[radius=2pt];
        \filldraw (bl) circle[radius=2pt];
        \filldraw (br) circle[radius=2pt];
  \draw (tl) to[out=-45, in=225] (tr);
  \draw (tl) -- (tr);
 \draw (bl) to[out=45, in=135] (br);
 \draw (bl) -- (br);
    \end{tikzpicture} + \begin{tikzpicture}[scale=0.5,baseline=(vert_cent.base)]
	\node (vert_cent) at (0,0) {$\phantom{\cdot}$};
	\node (center) at (0,0) {};
	\def\radius{1cm};
	\node[inner sep=0pt] (tr) at (45:\radius) {};
	\node[inner sep=0pt] (tl) at (135:\radius) {};
	\node[inner sep=0pt] (bl) at (225:\radius) {};
        \node[inner sep=0pt] (br) at (315:\radius) {};
	\draw (center) circle[radius=\radius];
	\filldraw (tr) circle[radius=2pt];
        \filldraw (tl) circle[radius=2pt];
        \filldraw (bl) circle[radius=2pt];
        \filldraw (br) circle[radius=2pt];
	\draw (tr) -- (bl);
	\draw (tl) -- (br);
 \draw (tl) to[out=-45, in=225] (tr);
 \draw (bl) to[out=45, in=135] (br);
	\end{tikzpicture}  \right) \, ,
\end{equation}
in which to each vertex corresponds an integration in real space and symmetry coefficient have been taken into account because now we are working with proper Feynman diagrams. Nevertheless, the tensor contraction of $\lambda$ multiplying the diagrams can be deduced in analogy with the previous section with the only difference that $\lambda$
appearing in Eq.~\eqref{eq:F-general} is the bare coupling.

\subsection{$F$-diagrams at leading and next-to-leading order}\label{sect:1loop-F}

The diagrams in Eqs.~\eqref{eq:afun0} and \eqref{eq:afun1} can be computed analytically and in arbitrary dimension $d$ using two well-known integrals on the sphere \cite{Fei:2015oha,Cardy:1988cwa}. We have 
\begin{equation*}\label{eq:A0F,A1F}
\begin{split}
  & \begin{tikzpicture}[scale=0.5,baseline=(vert_cent.base)]
	\node (vert_cent) at (0,0) {$\phantom{\cdot}$};
	\node (center) at (0,0) {};
	\def\radius{1cm};
	\node[inner sep=0pt] (l) at (0:\radius) {};
    \node[inner sep=0pt] (l1) at (0:1.5\radius) {$y$};
	\node[inner sep=0pt] (r) at (180:\radius) {};
    \node[inner sep=0pt] (r1) at (180:1.5\radius) {$x$};
	\draw (center) circle[radius=\radius];
	\filldraw (r) circle[radius=2pt];
    \filldraw (l) circle[radius=2pt];
	\draw (r) to[out=45, in=135] (l);
 \draw (r) to[out=-45, in=-135] (l);
	\end{tikzpicture}    
 = \mathcal{N}_0 \int {\rm d}^dx \sqrt{g_x} \int {\rm d}^dy \sqrt{g_y} \ \frac{C_{d,s}^4}{\sigma(x,y)^{4(d-s)}} \\
 & \qquad = \mathcal{N}_0 \ C_{d,s}^4 (2R)^{2(d-2(d-s))}\frac{2^{1-d}\pi^{d+\frac{1}{2}}\Gamma(\frac{d}{2}-2(d-s))}{\Gamma(\frac{d+1}{2})\Gamma(d-2(d-s))} \quad \xrightarrow{s\to \frac{d+\varepsilon}{2}} \\
  &\mathcal{N}_0 \frac{2^{1-3d}\pi^{1/2-d}\Gamma(-d/2)}{\Gamma\left(\frac{1+d}{2}\right)} \varepsilon (1+\varepsilon(\gamma+2\log(R)+\psi(-d/2)-2\psi(d/4))) + \mathcal{O}(\varepsilon^3)
\, , 
\end{split}
\end{equation*}
where $R$ is the radius of $S^d$ and $\gamma\simeq 0.5772$ is the Euler-Mascheroni constant, and
\begin{equation*}
\begin{split}
  & \begin{tikzpicture}[scale=0.5,baseline=(vert_cent.base)]
	\node (vert_cent) at (0,0) {$\phantom{\cdot}$};
	\node (center) at (0,0) {};
	\def\radius{1cm};
	\node[inner sep=0pt] (tr) at (0:\radius) {};
    \node[inner sep=0pt] (tr1) at (0:1.5\radius) {$y$};
	\node[inner sep=0pt] (tl) at (120:\radius) {};
    \node[inner sep=0pt] (tl1) at (120:1.5\radius) {$x$};
	\node[inner sep=0pt] (bl) at (240:\radius) {};
    \node[inner sep=0pt] (bl1) at (240:1.5\radius) {$z$};
	\draw (center) circle[radius=\radius];
	\filldraw (tr) circle[radius=2pt];
        \filldraw (tl) circle[radius=2pt];
        \filldraw (bl) circle[radius=2pt];
	\draw (tr) to[out=180, in=-60] (tl);
	\draw (tl) to[out=-60, in=60] (bl);
 \draw (tr) to[out=180, in=60] (bl);
	\end{tikzpicture}
 \, 
   = \mathcal{N}_1 \int {\rm d}^dx \sqrt{g_x} \int {\rm d}^dy \sqrt{g_y} \int {\rm d}^dz \sqrt{g_z} \ \frac{C_{d,s}^6}{\left[\sigma(x,y)\sigma(x,z)\sigma(z,y)\right]^{2(d-s)}} \\
  & \qquad = \mathcal{N}_1 \ C_{d,s}^6 R^{3(d-2(d-s))}\frac{8\pi^{\frac{3}{2}(1+d)}\Gamma\left(d-\frac{6(d-s)}{2}\right)}{\Gamma(d)\Gamma\left(\frac{1+d-2(d-s)}{2}\right)^3} \quad  \xrightarrow{s\to \frac{d+\varepsilon}{2}} \\
 & \mathcal{N}_1  \frac{8^{1-d}\pi^{-3d/2}\Gamma(-d/2)}{\Gamma(d)}\left( 1- \frac{3}{2}\varepsilon (\log(4)-2 \log(R)+\psi(1/2)-\psi(-d/2)+2\psi(d/4))\right) +\mathcal{O}(\varepsilon^2)
 \, .
 \end{split}
\end{equation*}
We have denoted with $\mathcal{N}_0$ and $\mathcal{N}_1$ the overall tensor contractions inclusive of their symmetry coefficients,
\begin{equation}
    \mathcal{N}_0=4! \ (\lambda_{ijkl}\lambda_{ijkl})
    \,, \qquad
    \mathcal{N}_1= 3 (4!)^2 \ (\lambda_{ijkl}\lambda_{klmn}\lambda_{mnij}) \, .
\end{equation}

\subsection{Next-to-next-to-leading $F$-diagrams}\label{sect:2loop-F}

The computation of the next diagrams is more involved and it is explicitly discussed in Appendix~\ref{sect:MB}, where the $d=2$ case of Eq.~\eqref{eq:A2Fb} is derived as an example. Only the final results of the calculations are reported here. 
We find
\begin{equation}\label{eq:A2Fb}
 \begin{split}
 \begin{tikzpicture}[scale=0.5,baseline=(vert_cent.base)]
	\node (vert_cent) at (0,0) {$\phantom{\cdot}$};
	\node (center) at (0,0) {};
	\def\radius{1cm};
	\node[inner sep=0pt] (tr) at (45:\radius) {};
	\node[inner sep=0pt] (tl) at (135:\radius) {};
	\node[inner sep=0pt] (bl) at (225:\radius) {};
        \node[inner sep=0pt] (br) at (315:\radius) {};
	\draw (center) circle[radius=\radius];
	\filldraw (tr) circle[radius=2pt];
        \filldraw (tl) circle[radius=2pt];
        \filldraw (bl) circle[radius=2pt];
        \filldraw (br) circle[radius=2pt];
	\draw (tr) to[out=225, in=135] (br);
	\draw (tl) to[out=-45, in=45] (bl);
  \draw (tl) to[out=-45, in=225] (tr);
 \draw (bl) to[out=45, in=135] (br);
    \end{tikzpicture}
 \, 
= \mathcal{N}_{21} (2R)^{4\varepsilon}
\left\{
\begin{array}{ll}
-\frac{20 \pi ^4}{\varepsilon ^2}-\frac{40 \pi ^4}{\varepsilon }-80 \pi ^4 + O(\varepsilon) & \quad {\rm in}~~d=2 \\
\frac{160 \pi ^6}{3 \varepsilon }-\frac{320}{9} \pi ^6 (\log (64)-8) + O(\varepsilon) & \quad {\rm in}~~d=3 
\end{array}
\right.
\,,
\end{split}
\end{equation}
then
\begin{equation}\label{eq:A2Fc}
 \begin{split}
\begin{tikzpicture}[scale=0.5,baseline=(vert_cent.base)]
	\node (vert_cent) at (0,0) {$\phantom{\cdot}$};
	\node (center) at (0,0) {};
	\def\radius{1cm};
	\node[inner sep=0pt] (tr) at (45:\radius) {};
	\node[inner sep=0pt] (tl) at (135:\radius) {};
	\node[inner sep=0pt] (bl) at (225:\radius) {};
        \node[inner sep=0pt] (br) at (315:\radius) {};
	\draw (center) circle[radius=\radius];
	\filldraw (tr) circle[radius=2pt];
        \filldraw (tl) circle[radius=2pt];
        \filldraw (bl) circle[radius=2pt];
        \filldraw (br) circle[radius=2pt];
  \draw (tl) to[out=-45, in=225] (tr);
  \draw (tl) -- (tr);
 \draw (bl) to[out=45, in=135] (br);
 \draw (bl) -- (br);
    \end{tikzpicture}
 \, 
  = \mathcal{N}_{22} (2R)^{4\varepsilon}
\left\{
\begin{array}{ll}
4 \pi ^4 + O(\varepsilon) & \quad {\rm in}~~d=2 \\
O(\varepsilon) & \quad {\rm in}~~d=3 
\end{array}
\right. \,,
\end{split}
\end{equation}
and finally
\begin{equation}\label{eq:A2Fa}
 \begin{split}
\begin{tikzpicture}[scale=0.5,baseline=(vert_cent.base)]
	\node (vert_cent) at (0,0) {$\phantom{\cdot}$};
	\node (center) at (0,0) {};
	\def\radius{1cm};
	\node[inner sep=0pt] (tr) at (45:\radius) {};
	\node[inner sep=0pt] (tl) at (135:\radius) {};
	\node[inner sep=0pt] (bl) at (225:\radius) {};
        \node[inner sep=0pt] (br) at (315:\radius) {};
	\draw (center) circle[radius=\radius];
	\filldraw (tr) circle[radius=2pt];
        \filldraw (tl) circle[radius=2pt];
        \filldraw (bl) circle[radius=2pt];
        \filldraw (br) circle[radius=2pt];
	\draw (tr) -- (bl);
	\draw (tl) -- (br);
 \draw (tl) to[out=-45, in=225] (tr);
 \draw (bl) to[out=45, in=135] (br);
	\end{tikzpicture}
 \, 
= \mathcal{N}_{23} (2R)^{4\varepsilon}
\left\{
\begin{array}{ll}
-\frac{10 \pi ^4}{\varepsilon ^2}-\frac{2 \pi ^4 (10+\log (64))}{\varepsilon }-\pi ^4 \left(40+\log ^2(4)+12 \log (4)\right) + O(\varepsilon) & \quad {\rm in}~~d=2 \\
\frac{80 \pi ^6}{3 \varepsilon }-\frac{8}{9} \pi ^6 (-178+9 \pi +84 \log (2)) + O(\varepsilon) & \quad {\rm in}~~d=3 
\end{array}
\right. \,,
\end{split}
\end{equation}
in which we have also introduced shorthands for the tensor contractions with symmetry factors,
\begin{equation*}
\begin{split}
    &\mathcal{N}_{21}= 12 \ (4!)^2 \ 9 \ (\lambda_{ijkl} \lambda_{ijnp} \lambda_{klrs} \lambda_{nprs}) \, , \qquad  \mathcal{N}_{22}= 12 \ (4!)^2 \ 8 \ (\lambda_{ijnk} \lambda_{ijnp} \lambda_{klrs} \lambda_{plrs}) \ , \\
    &  \mathcal{N}_{23}= 12 \ (4!)^2 \ 36 \ (\lambda_{ijkl} \lambda_{ijnp} \lambda_{nkrs} \lambda_{plrs}) \, .
\end{split}
\end{equation*}

\subsection{Renormalized sphere free-energy}\label{sect:SphereFreeEnergy}

By combining together all the integrals in Eq.~\eqref{eq:F-general} we have an expression for the bare free-energy on the sphere. In order to obtain the renormalized expression, we need to express the bare couplings in the contractions $\mathcal{N}_i$s in terms of the renormalized ones.

For the relation we follow a notation that is close to Ref.~\cite{Benedetti:2020rrq}
\begin{equation}\label{eq:tensor-CT}
    \mu^{-\varepsilon}\lambda^0_{ijkl}=\lambda_{ijkl}+\frac{1}{2}D\left[ {\cal S}_3 \,
  \begin{tikzpicture}[scale=0.5,baseline=(vert_cent.base)]
	\node (vert_cent) at (0,0) {$\phantom{\cdot}$};
 \node (center) at (0,0) {};
	\def\radius{1cm};
	\node[inner sep=0pt] (l) at (180:\radius) {};
        \node[inner sep=0pt] (etl) at (160:1.7*\radius) {};
        \node[inner sep=0pt] (ebl) at (200:1.7*\radius) {};
        \node[inner sep=0pt] (itr) at (20:1.7*\radius) {};
        \node[inner sep=0pt] (ibr) at (-20:1.7*\radius) {};
        \node[inner sep=0pt] (r) at (0:\radius) {};
	       \draw (center) circle[radius=\radius];
	       \filldraw (l) circle[radius=2pt];
        \filldraw (r) circle[radius=2pt];
	\draw (l) -- (etl);
	\draw (l) -- (ebl);
	\draw (r) -- (itr);
    \draw (r) -- (ibr);
  \end{tikzpicture} \right] + \frac{1}{2} (D^2-S) \left[ {\cal S}_6\,
  \begin{tikzpicture}[scale=0.5,baseline=(vert_cent.base)]
	\node (vert_cent) at (0,0) {$\phantom{\cdot}$};
	\node (center) at (0,0) {};
	\def\radius{1cm};
	\node[inner sep=0pt] (tl) at (120:\radius) {};
        \node[inner sep=0pt] (etl) at (120:1.7*\radius) {};
        \node[inner sep=0pt] (bl) at (240:\radius) {};
        \node[inner sep=0pt] (ebl) at (240:1.7*\radius) {};
	\node[inner sep=0pt] (r) at (0:\radius) {};
        \node[inner sep=0pt] (etr) at (30:1.7*\radius) {};
        \node[inner sep=0pt] (ebr) at (-30:1.7*\radius) {};
	\draw (center) circle[radius=\radius];
	\filldraw (tl) circle[radius=2pt];
        \filldraw (bl) circle[radius=2pt];
	\filldraw (r) circle[radius=2pt];
	\draw (tl) -- (etl);
	\draw (bl) -- (ebl);
	\draw (r) -- (etr);
 \draw (r) -- (ebr);
        \draw (tl) to[out=315, in=45] (bl);
  \end{tikzpicture} \right] + \frac{1}{4} D^2 \left[ {\cal S}_3\,
  \begin{tikzpicture}[scale=0.5,baseline=(vert_cent.base)]
	\node (vert_cent) at (0,0) {$\phantom{\cdot}$};
	\node (center) at (0,0) {};
	\def\radius{1cm};
 \node[inner sep=0pt] (l) at (0:-1*\radius) {};
 \node[inner sep=0pt] (m) at (0:0*\radius) {};
 \node[inner sep=0pt] (r) at (0:1*\radius) {};
 \node[inner sep=0pt] (bl) at (30:-1.7*\radius) {};
 \node[inner sep=0pt] (tl) at (-30:-1.7*\radius) {};
 \node[inner sep=0pt] (br) at (150:-1.7*\radius) {};
 \node[inner sep=0pt] (tr) at (-150:-1.7*\radius) {};
	\filldraw (l) circle[radius=2pt];
       \filldraw (m) circle[radius=2pt];
	\filldraw (r) circle[radius=2pt];
	\draw (l) to[out=45, in=135] (m);
 \draw (l) to[out=-45, in=-135] (m);
	\draw (m) to[out=45, in=135] (r);
	\draw (m) to[out=-45, in=-135] (r);
 \draw (l) -- (tl);
 \draw (l) -- (bl);
 \draw (r) -- (tr);
 \draw (r) -- (br);
  \end{tikzpicture} \ \right] \, ,
\end{equation}
where now the diagrams represent again tensor contractions as in Sect.~\ref{sect:consistency} and a different notation is used for bare and renormalized couplings.
The divergent coefficients $D$ and $S$ are expressed as Laurent series in $\varepsilon$. For the renormalization of Eq.~\eqref{eq:F-general} we need the two loop result
\begin{equation}\label{eq:dS-integrals}
    \begin{split}
        & D=\frac{2^{1-d} \pi ^{-d/2}}{\varepsilon \  \Gamma \left(\frac{d}{2}\right)}-\frac{2^{-d} \pi ^{-d/2} \left(\psi\left(\frac{d}{2}\right)+\gamma \right)}{\Gamma
   \left(\frac{d}{2}\right)} + \mathcal{O}(\varepsilon) \\
        & S=\frac{2^{1-2 d} \pi ^{-d}}{\varepsilon ^2 \ \Gamma \left(\frac{d}{2}\right)^2}-\frac{(4 \pi )^{-d} \left(2 \psi \left(\frac{d}{4}\right)+\psi
   \left(\frac{d}{2}\right)+3 \gamma \right)}{\varepsilon \ \Gamma \left(\frac{d}{2}\right)^2} +\mathcal{O}(\varepsilon^0)  \, ,
    \end{split}
\end{equation}
which is responsible for the beta function
\begin{equation}\label{eq:beta-tensor}
\begin{split}
    &\beta_{ijkl}=-\varepsilon \left[ {\cal S}_1\,
  \begin{tikzpicture}[scale=0.5,baseline=(vert_cent.base)]
	\node (vert_cent) at (0,0) {$\phantom{\cdot}$};
 \node (center) at (0,0) {};
	\def\radius{1cm};
	    \node[inner sep=0pt] (tl) at (150:1.5\radius) {};
        \node[inner sep=0pt] (bl) at (210:1.5\radius) {};
        \node[inner sep=0pt] (tr) at (30:1.5\radius) {};
        \node[inner sep=0pt] (br) at (330:1.5\radius) {};
	    \filldraw (center) circle[radius=2pt];
	\draw (tl) -- (br);
	\draw (bl) -- (tr);
  \end{tikzpicture} \right] + \frac{2^{1-d} \pi^{-d/2}}{\Gamma\left(\frac{d}{2}\right)} \left[ {\cal S}_3 \,
  \begin{tikzpicture}[scale=0.5,baseline=(vert_cent.base)]
	\node (vert_cent) at (0,0) {$\phantom{\cdot}$};
 \node (center) at (0,0) {};
	\def\radius{1cm};
	\node[inner sep=0pt] (l) at (180:\radius) {};
        \node[inner sep=0pt] (etl) at (160:1.7*\radius) {};
        \node[inner sep=0pt] (ebl) at (200:1.7*\radius) {};
        \node[inner sep=0pt] (itr) at (20:1.7*\radius) {};
        \node[inner sep=0pt] (ibr) at (-20:1.7*\radius) {};
        \node[inner sep=0pt] (r) at (0:\radius) {};
	       \draw (center) circle[radius=\radius];
	       \filldraw (l) circle[radius=2pt];
        \filldraw (r) circle[radius=2pt];
	\draw (l) -- (etl);
	\draw (l) -- (ebl);
	\draw (r) -- (itr);
    \draw (r) -- (ibr);
  \end{tikzpicture} \right] \\
  & + \left(\frac{2^{1-d} \pi^{-d/2}}{\Gamma\left(\frac{d}{2}\right)} \right)^2  \left(2 \psi\left(\frac{d}{4}\right)-\psi\left(\frac{d}{2}\right)+\gamma \right) \left[ {\cal S}_6\,
  \begin{tikzpicture}[scale=0.5,baseline=(vert_cent.base)]
	\node (vert_cent) at (0,0) {$\phantom{\cdot}$};
	\node (center) at (0,0) {};
	\def\radius{1cm};
	\node[inner sep=0pt] (tl) at (120:\radius) {};
        \node[inner sep=0pt] (etl) at (120:1.7*\radius) {};
        \node[inner sep=0pt] (bl) at (240:\radius) {};
        \node[inner sep=0pt] (ebl) at (240:1.7*\radius) {};
	\node[inner sep=0pt] (r) at (0:\radius) {};
        \node[inner sep=0pt] (etr) at (30:1.7*\radius) {};
        \node[inner sep=0pt] (ebr) at (-30:1.7*\radius) {};
	\draw (center) circle[radius=\radius];
	\filldraw (tl) circle[radius=2pt];
        \filldraw (bl) circle[radius=2pt];
	\filldraw (r) circle[radius=2pt];
	\draw (tl) -- (etl);
	\draw (bl) -- (ebl);
	\draw (r) -- (etr);
 \draw (r) -- (ebr);
        \draw (tl) to[out=315, in=45] (bl);
  \end{tikzpicture} \right]
  \end{split} \,,
\end{equation}
that was used in Sect.~\ref{sect:consistency} for the contributions up to two loops. 

Substituting \eqref{eq:tensor-CT} into \eqref{eq:F-general}, we get the renormalized sphere free energy in $d=3$
\begin{equation}\label{eq:Sphere-Free-Energy-d=3}
    \begin{split}
        \delta \tilde{F}_3=&-\frac{9 (6 \log (R)-3 \pi +6 \gamma +14+\log (64))}{1024 \pi ^6} \ \lambda_{ijkl} \lambda_{ijnp} \lambda_{klrs} \lambda_{nprs} \\
        & -\frac{9 (12 \log (R)-9 \pi +12 \gamma +34+24 \log (2))}{1024 \pi ^6} \ \lambda_{ijkl} \lambda_{ijnp} \lambda_{nkrs} \lambda_{plrs} \\
        & + \left(\frac{3 \varepsilon  (6 \log (R)-3 \pi +6 \gamma +11+\log (512))}{256 \pi ^4}+\frac{3}{256 \pi ^4} \right) \lambda_{ijkl}\lambda_{klmn}\lambda_{mnij}  \\
        & - \left( \frac{\varepsilon ^2 (6 \log (R)-3 \pi +6 \gamma +8+\log (4096))}{768 \pi ^2}+\frac{\varepsilon }{256 \pi ^2} \right) \lambda_{ijkl}\lambda_{ijkl} \, ,
    \end{split}
\end{equation}
which is finite as expected.
Similarly in even $d$, we can obtain the (topological) Weyl anomaly $a_d$ by considering the $\log(R)$ term of the sphere free-energy, $F\supset(-1)^{\frac{d}{2}}a_d \log(R)$. The anomaly is then related to the $\tilde{F}$-function via $\tilde{F}_d=\frac{\pi}{2}a_d$. For $d=2$ we find
\begin{equation}\label{eq:Sphere-Free-Energy-d=2}
    \begin{split}
        \delta a_2 &=-\frac{81 (1+2 \gamma +\log (16))}{128 \pi ^4}  \lambda_{ijkl} \lambda_{ijnp} \lambda_{klrs} \lambda_{nprs} 
        \\& 
        -\frac{81 (1+2 \gamma +\log (64))}{64 \pi ^4} \ \lambda_{ijkl} \lambda_{ijnp} \lambda_{nkrs} \lambda_{plrs} \\
        & + \left(\frac{27 \varepsilon  (1+2 \gamma +\log (16))}{32 \pi ^3}+\frac{9}{32 \pi ^3} \right) \lambda_{ijkl}\lambda_{klmn}\lambda_{mnij} 
        \\
        &- \left( \frac{3 \varepsilon ^2 (1+2 \gamma +\log (16))}{32 \pi ^2}+\frac{3 \varepsilon }{32 \pi ^2} \right) \lambda_{ijkl}\lambda_{ijkl} \, .
    \end{split}
\end{equation}
Also in $d=2$ the anomaly can then be related to the standard Zamolodchikov's $c$-function as $c=3a_2$.
Eq.~\eqref{eq:Sphere-Free-Energy-d=2} describes the variation of the conformal anomaly in $d=2$ but it is straightforward to relate it to $c$ or $\delta \tilde{F}_2$ through the appropriate rescaling.

As next step we specialize the beta function of Eq.~\eqref{eq:beta-tensor} and the free energy to two specific fixed point of interest, the vector $O(N)$ and the hypercubic $H_N$, by computing the fixed point value of the coupling and inserting it into the projected free energy.

\subsection{The $O(N)$ fixed point}\label{sect:O(N)}

To project onto the $O(N)$ symmetric fixed point we specialize to the invariant coupling
\begin{equation}\label{eq:O(N)-proj}
  \lambda_{ijkl}=\frac{\lambda}{3} (\delta_{ij}  \delta_{kl}+\delta_{ik}  \delta_{jl}+\delta_{il}  \delta_{jk}) \, ,
\end{equation}
for which the beta function \eqref{eq:beta-tensor} is
\begin{equation}\label{eq:beta-O(N)}
    \beta_{\lambda}=-\varepsilon \lambda + \left(\frac{2^{1-d} \pi^{-d/2}}{\Gamma\left(\frac{d}{2}\right)}\right) (8+N) \ \lambda^2 + \left( \frac{2^{1-d} \pi^{-d/2}}{\Gamma\left(\frac{d}{2}\right)} \right)^2 2(22+5N) \left( 2 \psi \left(\frac{d}{4}\right)-\psi\left(\frac{d}{2}\right)+\gamma \right) \ \lambda^3 + \mathcal{O}(\lambda^4) \, ,
\end{equation}
and the fixed point $\lambda_*$ becomes
\begin{equation}\label{eq:fixed-point-O(N)}
\begin{split}
    \lambda_*=\left(\frac{2^{1-d}\pi^{-d/2}}{\Gamma\left(\frac{d}{2}\right)}\right)^{-1} &\biggl( \frac{\varepsilon }{N+8} -\varepsilon^2 \ \frac{2 (5 N+22)  \left(2 \psi\left(\frac{d}{4}\right)-\psi \left(\frac{d}{2}\right)+\gamma \right)}{(N+8)^3} \\
   & + \varepsilon ^3 \ \frac{8 (5 N+22)^2  \left(2 \psi \left(\frac{d}{4}\right)-\psi \left(\frac{d}{2}\right)+\gamma \right)^2}{(N+8)^5} \biggr) + \mathcal{O}(\varepsilon^4) \, .
\end{split}
\end{equation}
Therefore \eqref{eq:Sphere-Free-Energy-d=3} evaluated at the fixed point in $d=3$ gives
\begin{equation}\label{eq:F3-O(N)}
   \delta\tilde{F}_3^{O(N)}= -\varepsilon ^3 \frac{\pi ^2 N (N+2)}{576 (N+8)^2} + \varepsilon ^4 \frac{\pi ^2 N (N+2) (5 N+22) (-2+\pi -4 \log (2))}{192 (N+8)^4} + \mathcal{O}(\varepsilon^5) \, ,
\end{equation}
while \eqref{eq:Sphere-Free-Energy-d=2} in $d=2$ gives
\begin{equation}\label{eq:F2-O(N)}
    \delta c^{O(N)}= - \varepsilon ^3 \frac{N (N+2)}{8 (N+8)^2} - \varepsilon ^4 \frac{3 N (N+2) (5 N+22) \log (2)}{2 (N+8)^4} + \mathcal{O}(\varepsilon^5) \, .
\end{equation}
Our results match those in \cite{Giombi:2024zrt}.

\subsection{The hypercubic fixed point}\label{sect:H(N)}

We choose to specialize our calculation to a second fixed point, which is useful later in Section \ref{sect:matching} to argue that that the matching between $A$ and $\tilde{F}$ is independent of the choice of fixed point.

We can repeat the procedure applied in Sect.~\ref{sect:O(N)} for the hypercubic fixed point with discrete symmetry group $H_N$.
The invariant coupling is
\begin{equation}\label{eq:H(N)-projector}
    \lambda_{ijkl}=g_c \, \delta_{ijkl} +\frac{g_d}{3} (\delta_{ij}  \delta_{kl}+\delta_{ik}  \delta_{jl}+\delta_{il}  \delta_{jk}) \, ,
\end{equation}
where $\delta_{ijkl}=1$ iff all indices are the same and otherwise is zero.
Inserting \eqref{eq:H(N)-projector} in the tensor beta function \eqref{eq:beta-tensor} gives us the beta functions
\begin{equation}\label{eq:beta-gc-gd}
\begin{split}
    &\beta_c=-\varepsilon g_c + \left(\frac{2^{1-d}\pi^{-d/2}}{\Gamma\left(\frac{d}{2}\right)}\right) g_c (3 g_c+4 g_d)
    + \left(\frac{2^{1-d}\pi^{-d/2}}{\Gamma\left(\frac{d}{2}\right)}\right)^2  \frac{2}{3} g_c \ \times \\
    & \times \left(2 \psi\left(\frac{d}{4}\right)-\psi\left(\frac{d}{2}\right)+\gamma \right) \left(9 g_c^2+24 g_c
   g_d+g_d^2 (N+14)\right) +\mathcal{O}\left((g_c g_d)^2\right) \, , \\
   & \beta_d= - \varepsilon g_d + \left(\frac{2^{1-d}\pi^{-d/2}}{\Gamma\left(\frac{d}{2}\right)}\right) g_d (6 g_c+g_d (N+8)) + \left(\frac{2^{1-d}\pi^{-d/2}}{\Gamma\left(\frac{d}{2}\right)}\right)^2 2 g_d \ \times \\
   & \times \left(2 \psi\left(\frac{d}{4}\right)-\psi\left(\frac{d}{2}\right)+\gamma \right) \left(9 g_c^2+36 g_c g_d+g_d^2 (5 N+22)\right) +\mathcal{O}\left((g_c g_d)^2\right) \, .
   \end{split}
\end{equation}
The system admits more than one fixed point, among which there is the one with hypercubic symmetry
\begin{equation}\label{eq:gc-gd-FP}
\begin{split}
   & g_c^*= \left(\frac{2^{1-d}\pi^{-d/2}}{\Gamma\left(\frac{d}{2}\right)}\right)^{-1}\biggl( \varepsilon \frac{(N-4) }{9 N}  -\varepsilon ^2 \frac{2 \left(N^3+5 N^2-30 N+24\right)  \left(2 \psi\left(\frac{d}{4}\right)-\psi\left(\frac{d}{2}\right)+\gamma \right)}{27 N^3} \\
    & \qquad+ \varepsilon ^3 \frac{8
   (N-1) \left(N^4+5 N^3+28 N^2-172 N+144\right)  \left(2 \psi \left(\frac{d}{4}\right)-\psi\left(\frac{d}{2}\right)+\gamma \right)^2}{81 N^5}\biggr) + \mathcal{O}(\varepsilon^4) \, , \\
   & g_d^*= \left(\frac{2^{1-d}\pi^{-d/2}}{\Gamma\left(\frac{d}{2}\right)}\right)^{-1} \biggl( \frac{\varepsilon }{3 N} + \varepsilon ^2 \frac{2 \left(N^2-7 N+6\right)  \left(2 \psi \left(\frac{d}{4}\right)-\psi \left(\frac{d}{2}\right)+\gamma \right)}{9 N^3} \\
   & \qquad - \varepsilon ^3 \frac{8 (N-1)
   \left(N^3+5 N^2-40 N+36\right)  \left(2 \psi \left(\frac{d}{4}\right)-\psi \left(\frac{d}{2}\right)+\gamma \right)^2}{27
   N^5} \biggr) + \mathcal{O}(\varepsilon^4) \, .
\end{split}
\end{equation}
Evaluating \eqref{eq:Sphere-Free-Energy-d=3} at the fixed point we find in $d=3$
\begin{equation}\label{eq:F3-H(N)}
   \delta \tilde{F}_3^{H_N}= -\varepsilon ^3 \frac{\pi ^2 \left(N^2+N-2\right) }{15552 N} + \varepsilon ^4 \frac{\pi ^2 (N-1)^2 \left(N^2+12\right)  (-2+\pi -4 \log (2))}{15552 N^3} + \mathcal{O}(\varepsilon^5) \, ,
\end{equation}
and in $d=2$
\begin{equation}\label{eq:F2-H(N)}
    \delta c^{H_N}= - \varepsilon ^3 \frac{\left(N^2+N-2\right) }{216 N}- \varepsilon ^4 \frac{(N-1)^2 \left(N^2+12\right) \log (2)}{54 N^3} + \mathcal{O}(\varepsilon^5) \, .
\end{equation}
In Section \ref{sect:matching} we will match the gradient structure both for the $O(N)$ and the hypercubic cases, to show that our procedure is independent of the particular fixed point of choice.

\section{Two-point function}\label{sect:two-point}

For the matching of the gradient structure of the RG flow we need the candidate metric on the theory space, which is the tensor structure coming from the two-point function of the marginal operators in the interacting theory, as suggested by Zamolodchikov's $c$-theorem \cite{Zamolodchikov:1986gt}. 
As discussed in Section \ref{sect:intro}, the marginal operators are $\phi^4_I$ with \emph{naive} dimension $2(d-s)\to d-\varepsilon$ for $s=\frac{d+\varepsilon}{2}$.
Our aim in this section is therefore to compute and renormalize the two-point function $\langle \phi_I^4(x) \phi_J^4(y) \rangle$, which we do with the Feynman diagrams diagrammatically respresented by Eqs.~\eqref{eq:metric0} and \eqref{eq:metric1}.
For consistency with Sect.~\ref{sect:F-function} we do it on $S^d$, although it is not strictly necessary.

The $0$-loops contribution to the metric is represented by \eqref{eq:metric0}. The computation is trivial because it is simply a product of free scalar propagators given in \eqref{eq:propagator}. At the order that we need, we find
\begin{equation}\label{eq:2PF-0loop}
   \left[ \, \begin{tikzpicture}[scale=0.5,baseline=(vert_cent.base)]
	\node (vert_cent) at (0,0) {$\phantom{\cdot}$};
	\node (center) at (0,0) {};
	\def\vlength{0.7cm};
        \def\hlength{1cm};
        \node[inner sep=0pt] (l1) at (-1*\hlength,1.25*\vlength) {};
        \node[inner sep=0pt] (l2) at (-1*\hlength,0.5*\vlength) {};
	\node[inner sep=0pt] (l3) at (-1*\hlength,-0.25*\vlength) {};
    \node[inner sep=0pt] (l3tmp) at (-1.5*\hlength,0*\vlength) {$x$};
 	\node[inner sep=0pt] (l4) at (-1*\hlength,-1*\vlength) {};
	\node[inner sep=0pt] (r1) at (1*\hlength,1.25*\vlength) {};
	\node[inner sep=0pt] (r2) at (1*\hlength,0.5*\vlength){};
        \node[inner sep=0pt] (r3) at (1*\hlength,-0.25*\vlength) {};
        \node[inner sep=0pt] (r3tmp) at (1.5*\hlength,0*\vlength) {$y$};
        \node[inner sep=0pt] (r4) at (1*\hlength,-1*\vlength) {};
        \draw (l1) -- (r1);
        \draw (l2) -- (r2);
        \draw (l3) -- (r3);
        \draw (l4) -- (r4);
    \end{tikzpicture} \, \right] = \mathcal{M}^0_{IJ} \frac{C_{d,s}^4}{\sigma(x,y)^{4(d-s)}}= \mathcal{M}^0_{IJ} \left\{ \begin{array}{lr}
         \frac{1}{16 \pi^4 (x-y)^4} + O(\varepsilon) \quad \text{in} \, d=2 \, , \\
         \frac{1}{64 \pi^6 (x-y)^6} + O(\varepsilon) \quad \text{in} \, d=3 \, ,
    \end{array} \right.
\end{equation}
where $\mathcal{M}^0_{IJ}= \delta^{ijkl}_{mnpr}$ indicates the identity in the space of symmetric tensors with indices $I=\{ ijkl\}$, $J=\{mnpr\}$ generalized indices.

The $1$-loop result is the first nontrivial contribution to the metric and fortunately the only one that we need for the matching.
The calculation of the diagram in \eqref{eq:metric1}
requires integrating over the intermediate $z$ space-time point in \eqref{eq:2PF-1loop} on $S^d$, which can be done with the methods illustrated in Appendix \ref{sect:MB}:
\begin{equation}\label{eq:2PF-1loop}
\begin{split}
   & \left[ \ \begin{tikzpicture}[scale=0.5,baseline=(vert_cent.base)]
	\node (vert_cent) at (0,-0.5) {$\phantom{\cdot}$};
 \node (center) at (0,0) {};
	\def\radius{0.7cm};
 	\def\vlength{0.5cm};
        \def\hlength{1cm};
	\node[inner sep=0pt] (tl) at (135:0) {};
    \node[inner sep=0pt] (tltmp) at (90:0.7*\radius) {$z$};
        \node[inner sep=0pt] (etl) at (160:1.7*\radius) {};
        \node[inner sep=0pt] (bl) at (225:0) {};
        \node[inner sep=0pt] (ebl) at (200:1.7*\radius) {};
        \node[inner sep=0pt] (tr) at (20:1.7*\radius) {};
        \node[inner sep=0pt] (br) at (-20:1.7*\radius) {};
        \node[inner sep=0pt] (itr) at (45:0) {};
        \node[inner sep=0pt] (ibr) at (-45:0) {};
        \node[inner sep=0pt] (l3) at (-1.*\hlength,-1.5*\vlength) {};
        \node[inner sep=0pt] (l3tmp) at (-1.5*\hlength,-1*\vlength) {$x$};
 	      \node[inner sep=0pt] (l4) at (-1*\hlength,-2.5*\vlength) {};
        \node[inner sep=0pt] (r3) at (1*\hlength,-1.5*\vlength) {};
        \node[inner sep=0pt] (r3tmp) at (1.5*\hlength,-1*\vlength) {$y$};
        \node[inner sep=0pt] (r4) at (1*\hlength,-2.5*\vlength) {};
	       \filldraw (tl) circle[radius=2pt];
        \filldraw (bl) circle[radius=2pt];
        \filldraw (itr) circle[radius=2pt];
        \filldraw (ibr) circle[radius=2pt];
	\draw (tl) -- (etl);
	\draw (bl) -- (ebl);
	\draw (itr) -- (tr);
    \draw (ibr) -- (br);
    \draw (l3) -- (r3);
    \draw (l4) -- (r4);
  \end{tikzpicture} \ \right]
    = \mathcal{M}^1_{IJ} \frac{C_{d,s}^6}{\sigma(x,y)^{2(d-s)}} \int {\rm d}^d z \sqrt{g_z} \frac{1}{\left[ \sigma(x,z)\sigma(z,y) \right]^{2(d-s)}}  \\
  & = \mathcal{M}^1_{IJ} \left\{ \begin{array}{ll}
         \frac{1}{16 \pi ^5 \varepsilon  (x-y)^4} + \frac{3 (-\log (x-y)+\gamma +\log (2))}{16 \pi ^5 (x-y)^4} + O(\varepsilon) \qquad & \text{in} \quad d=2 \,, \\
         \frac{1}{64 \pi ^8 \varepsilon  (x-y)^6}-\frac{3 \left(-\log (x-y)+\log (2)+\psi\left(\frac{3}{4}\right)\right)}{64 \pi ^8 (x-y)^6} + O(\varepsilon) \qquad & \text{in} \quad d=3 \, ,
  \end{array} \right. 
\end{split}
\end{equation}
where the tensor coefficient is defined as $\mathcal{M}^1_{IJ}=\lambda_{ijpr}\delta_{km}\delta_{ln} + \cdots $ and includes all necessary permutations of the indices.

Eqs.~\eqref{eq:2PF-0loop} and \eqref{eq:2PF-1loop} are the bare two-point function $\langle \phi_I^4(x)\phi_I^4(y) \rangle$, which we need to renormalize first and then write in a basis of scaling eigenoperators to identify it with the proposed metric of Zamolodchikov.

\subsection{Renormalizing the two-point function at the first nontrivial order}\label{sect:2PF-renorm}

This type of diagonalization and matching of the correlator follows closely the case of the short range $\phi^4$ theory given in Ref.~\cite{Pannell:2025ixz}. Even if the complete theory \eqref{eq:action} of this paper is nonlocal, the interaction term is. It is thus possible to carry out the renormalization of the correlators in the usual way, subtracting the poles through local counterterms.

We renormalize the diagram of Eq.~\eqref{eq:2PF-1loop}) by subtracting the $1/\varepsilon$ poles with counterterms.
We choose to represent the contribution of the counterterms as
\begin{equation}\label{eq:counterterms}
    \begin{tikzpicture}[scale=0.5,baseline=(vert_cent.base)]
	\node[inner sep=0pt] (vert_cent) at (0,0) {$\phantom{\cdot}$};
	\node[inner sep=0pt] (center) at (0,0) {};
    \node[inner sep=0pt] (left) at (-1.5cm,0) {};
    \node[inner sep=0pt] (right) at (1.5cm,0) {};
	\def\vlength{0.7cm};
        \def\hlength{1cm};
        \node[inner sep=0pt] (l1) at (-1*\hlength,1.25*\vlength) {};
        \node[inner sep=0pt] (l2) at (-1*\hlength,0.5*\vlength) {};
	\node[inner sep=0pt] (l3) at (-1*\hlength,-0.25*\vlength) {};
    \node[inner sep=0pt] (l3tmp) at (-2.7*\hlength,0*\vlength) {$x$};
 	\node[inner sep=0pt] (l4) at (-1*\hlength,-1*\vlength) {};
	\node[inner sep=0pt] (r1) at (1*\hlength,1.25*\vlength) {};
	\node[inner sep=0pt] (r2) at (1*\hlength,0.5*\vlength){};
        \node[inner sep=0pt] (r3) at (1*\hlength,-0.25*\vlength) {};
        \node[inner sep=0pt] (r3tmp) at (2.7*\hlength,0*\vlength) {$y$};
        \node[inner sep=0pt] (r4) at (1*\hlength,-1*\vlength) {};
        \draw (left) to[out=90, in=90] (right);
        \draw (left) to[out=20, in=160] (right);
        \draw (left) to[out=340, in=200] (right);
        \draw (left) to[out=270, in=270] (right);
        \filldraw[fill=white] (left) circle[radius=20pt];
        \filldraw[fill=white] (right) circle[radius=20pt];
        \node[inner sep=0pt] (leftw) at (-1.5cm,0) {$\delta\phi^4$};
        \node[inner sep=0pt] (rightw) at (1.5cm,0) {$\phi^4$};
    \end{tikzpicture} 
    + 
        \begin{tikzpicture}[scale=0.5,baseline=(vert_cent.base)]
	\node[inner sep=0pt] (vert_cent) at (0,0) {$\phantom{\cdot}$};
	\node[inner sep=0pt] (center) at (0,0) {};
    \node[inner sep=0pt] (left) at (-1.5cm,0) {};
    \node[inner sep=0pt] (right) at (1.5cm,0) {};
	\def\vlength{0.7cm};
        \def\hlength{1cm};
        \node[inner sep=0pt] (l1) at (-1*\hlength,1.25*\vlength) {};
        \node[inner sep=0pt] (l2) at (-1*\hlength,0.5*\vlength) {};
	\node[inner sep=0pt] (l3) at (-1*\hlength,-0.25*\vlength) {};
    \node[inner sep=0pt] (l3tmp) at (-2.7*\hlength,0*\vlength) {$x$};
 	\node[inner sep=0pt] (l4) at (-1*\hlength,-1*\vlength) {};
	\node[inner sep=0pt] (r1) at (1*\hlength,1.25*\vlength) {};
	\node[inner sep=0pt] (r2) at (1*\hlength,0.5*\vlength){};
        \node[inner sep=0pt] (r3) at (1*\hlength,-0.25*\vlength) {};
        \node[inner sep=0pt] (r3tmp) at (2.7*\hlength,0*\vlength) {$y$};
        \node[inner sep=0pt] (r4) at (1*\hlength,-1*\vlength) {};
        \draw (left) to[out=90, in=90] (right);
        \draw (left) to[out=20, in=160] (right);
        \draw (left) to[out=340, in=200] (right);
        \draw (left) to[out=270, in=270] (right);
        \filldraw[fill=white] (left) circle[radius=20pt];
        \filldraw[fill=white] (right) circle[radius=20pt];
        \node[inner sep=0pt] (leftw) at (-1.5cm,0) {$\phi^4$};
        \node[inner sep=0pt] (rightw) at (1.5cm,0) {$\delta\phi^4$};
    \end{tikzpicture} = \frac{1}{\varepsilon} \frac{2 (4!) C_{d,s}^4 \delta Z_{IJ}}{\sigma(x,y)^{4(d-s)}} \ ,
\end{equation}
where the $\delta \phi^4$ circle represents the counterterm to $\phi^4$ and the $\phi^4$ circle represents the bare operator itself. Notice that this graphical representation is slightly different from the one of the rest of the paper, but makes it easier to represent the local insertion of the counterterm. Here we are assuming that no important mixing occurs between the bare $\phi^4$ operator and other lower order operators, which can be verified in analogy to the short-range model.
The renormalization $\delta Z_{IJ}$ is determined by requiring that the pole cancels with \eqref{eq:2PF-1loop}, which gives
\begin{equation}\label{eq:deltaZeta}
    \delta Z_{IJ}= \delta Z  \, \mathcal{M}^1_{IJ} \quad {\rm with} \quad \delta Z= \frac{1}{48 \, \pi^{d-1}} \, . 
\end{equation}

As a consistency check we should first verify that $\delta Z_{IJ} \propto \partial_I \beta^J$ at order $\mathcal{O}(\lambda,\varepsilon^0)$, in agreement with the vertex renormalization of the original action.
We start with the beta function at order $\lambda^2$ from \eqref{eq:beta1} with the coefficient $b_1$ from \eqref{eq:beta1-values} and multiply it by the factor $\frac{1}{(4\pi)^{d/2}\Gamma(d/2)}$ to account for the rescaling in Ref.~\cite{Benedetti:2020rrq} given in Eq.~\eqref{eq:rescaling-benedetti}. Then taking a derivative with respect to the coupling $\lambda^I$ we get
\begin{equation}
    \partial_I \beta^{(1)}_J=\frac{1}{12}\frac{1}{(4\pi)^{d/2}\Gamma(d/2)} \mathcal{M}^1_{IJ} \, \longrightarrow \, \mathcal{M}^1_{IJ} \left\{ 
    \begin{array}{lr}
      \frac{1}{48 \pi} \quad & \text{in} \, d=2 \,, \\
         \frac{1}{48 \pi^2} \quad & \text{in} \, d=3 \, .
    \end{array} \right.
\end{equation}
which agrees with \eqref{eq:deltaZeta} (the arrow represent the rescaling of the coupling).

We now need to change to the basis of operators $[O]^\alpha$ that diagonalize scale transformations and appear in Zamolodchikov's metric. Their form is
\begin{equation}
    [O]^\alpha(x)=v_I^\alpha \phi_I^4(x) \, ,
\end{equation}
where $v_I^\alpha$ is an eigenvector of the mixing matrix $S_{IJ}$, i.e.\
\begin{equation}
    S_{IJ}v_{J}^\alpha=-\varepsilon \ (1-\gamma^\alpha)v_I^\alpha \ ,
\end{equation}
and the mixing matrix can be computed from the stability matrix $\partial\beta \sim {\cal M}^{1}_{IJ}$ as $S_{IJ}=-\varepsilon \delta_{IJ} - \delta Z \, {\cal M}^{1}_{IJ}$ (we have already established the relation of $\partial\beta$ with ${\cal M}^{1}$).
With the change of basis the two point function of the scaling operators becomes
\begin{equation}\label{2PF-O(x)O(y)}
    \langle [O]^\alpha(x)[O]^\beta(y) \rangle=\langle \phi_I^4(x) \phi_J^4(y) \rangle v_I^\alpha v_J^\beta \equiv G_{IJ} \, v_I^\alpha v_J^\beta \, ,
\end{equation}
where $G_{IJ}\equiv\langle \phi_I^4(x) \phi_J^4(y) \rangle$ to order $\mathcal{O}(\lambda^2)$ is the sum of equations \eqref{eq:2PF-0loop}, \eqref{eq:2PF-1loop} and \eqref{eq:counterterms}.

Now we assume that Eq.~\eqref{2PF-O(x)O(y)} must also be equal to the expression for a two-point function in a CFT,
\begin{equation}\label{eq:2PF-CFT}
    \langle [O]^\alpha(x)[O]^\beta(y) \rangle=\frac{C_0^\alpha + \varepsilon \ C_1^\alpha}{\sigma(x,y)^{4(d-s)+2 \varepsilon \gamma^\alpha}}\delta_{\alpha \beta} \, ,
\end{equation}
where we take into account that the operators may not be normalized in the usual way of primary CFT operators. The comparison of \eqref{2PF-O(x)O(y)} and \eqref{eq:2PF-CFT}
allows to determine $C_1^\alpha=-\gamma^\alpha C_0^\alpha=\frac{1}{(2 \pi)^{2d}}$, while $\gamma^\alpha$ is in general some complicated object that depends on the particular fixed point under study.

In the basis of renormalized scaling operators we have
\begin{equation}\label{eq:G-alpha-beta}
    C_{\alpha\beta}=G_{IJ} \, v_I^\alpha v_J^\beta = \left( m_0 + 2 \varepsilon \delta Z \gamma^\alpha m_1 \right) \delta_{\alpha \beta} =1+\varepsilon \, \alpha_0 \left( 1 + 2 \delta Z \gamma^\alpha\frac{\alpha_2}{\alpha_0} \right) \delta_{\alpha \beta} \, ,
\end{equation}
where $\varepsilon \delta Z$ is finite and given in Eq.~\eqref{eq:deltaZeta}, while $m_0=1+\varepsilon \, \alpha_0 +\mathcal{O}(\varepsilon^2)$ and $m_1=\alpha_2 + \mathcal{O}(\varepsilon)$ are the components of the metric from equations \eqref{eq:metric0} and \eqref{eq:metric1}.
The simple consequence is that \eqref{eq:G-alpha-beta} is the tensor which generalizes Zamolodchikov's metric and that it could be matched with the RG counterpart of Sect.~\ref{sect:consistency} by appropriately choosing some of undetermined parameters.

\section{Matching the gradient flow structures}\label{sect:matching}

We can now match together the $\tilde{F}$-function and the metric $C_{\alpha \beta}$ computed via CFT methods with their counterparts $A$ and $G_{IJ}$ of the RG framework of Sect.~\ref{sect:consistency}. We begin with the $O(N)$ model.

First, we rewrite the $A$-function using \eqref{eq:afun0}, \eqref{eq:afun1}, \eqref{eq:afun2} and the gradient solutions from Sect.~\ref{sect:consistency}. Then we evaluate it onto the $O(N)$ fixed point using the coupling in \eqref{eq:O(N)-proj} and the value for the fixed point in \eqref{eq:fixed-point-O(N)}. The $A$-function becomes
\begin{equation}\label{eq:AO(N)}
\begin{split}
   A^{O(N)}_d=& \left(\frac{2^{1-d}\pi^{-d/2}}{\Gamma\left(\frac{d}{2}\right)}\right)^{-2} \biggl( -\varepsilon ^3\frac{N (N+2) }{18 (N+8)^2} - \varepsilon^4 \frac{\alpha_0 N (N+2)}{18 (N+8)^2} \\
  & + \varepsilon^4 \frac{N (N+2) (5 N+22) \left(6 \left(\psi\left(\frac{d}{4}\right)+\gamma \right)-3 \left(\psi \left(\frac{d}{2}\right)+\gamma \right)\right)}{18
   (N+8)^4} \biggr)  \\
  & \qquad - \left(\frac{2^{1-d}\pi^{-d/2}}{\Gamma\left(\frac{d}{2}\right)}\right)^{-3}   \varepsilon ^4 \frac{\alpha_2 N (N+2)}{324 (N+8)^2} + \mathcal{O}(\varepsilon^5) \, ,
\end{split}
\end{equation}
with the undetermined $\alpha$ parameters.
This expression can be matched in $d=2,3$ with equations \eqref{eq:F3-O(N)} and $\frac{\pi}{6}$ times \eqref{eq:F2-O(N)}, since $\tilde{F}_2 = \frac{\pi}{2} a_2=\frac{\pi}{6}  c$, at the $\varepsilon^4$ order by rescaling \eqref{eq:AO(N)} up to an overall function of the dimension $\Omega(d)$ and choosing a value of $\alpha_2$ that is linearly related to $\alpha_0$:
\begin{equation}
    \delta \tilde{F}^{O(N)}_{d} = \Omega(d) \ A^{O(N)}_d(\alpha_2(\alpha_0)) \, ,
\end{equation}
such that
\begin{equation}\label{eq:matched-coefficients}
    \Omega(d)=\left\{ \begin{array}{ll}
         \frac{3}{32 \, \pi} \quad &\text{in} \quad d=2 \,,  \\
         \frac{1}{128 \, \pi^2} \quad &\text{in} \quad d=3 \,,
    \end{array} \right.  \qquad \alpha_2=-\frac{9}{\pi^{d-1}}\alpha_0 \, ,
\end{equation}
which we can generalize to arbitrary $d$ as\footnote{We are grateful to L.~Fraser-Taliente for pointing out this expression.}
\begin{equation}
    \Omega(d)=\frac{3}{2^{2d}\pi^{d-1}\Gamma(d+1)} \, .
\end{equation}
In short, for a specific choice of the $\alpha$s we have that $A$ and $\tilde{F}$ are the same function of $\varepsilon$ up to an overall constant. Notice that the parameter $\alpha_0$ is still undetermined by the matching.

In order to strengthen the conclusion, we then repeat the same procedure but with the quantities evaluated at the hypercubic fixed point.
Using the hypercubic projector in \eqref{eq:H(N)-projector} and the fixed point values in \eqref{eq:gc-gd-FP}, the $A$ function becomes
\begin{equation}
\begin{split}
    A^{H(N)}_d= & - \left(\frac{2^{1-d}\pi^{-d/2}}{\Gamma\left(\frac{d}{2}\right)}\right)^{-2} \biggl( \varepsilon ^3\frac{\left(N^2+N-2\right)}{486 N} + \varepsilon^4 \frac{\alpha_0 (N-1) (N+2)}{486 N} \\
    & \quad - \varepsilon^4 \frac{(N-1)^2 \left(N^2+12\right) \left(2 \left(\psi \left(\frac{d}{4}\right)+\gamma \right)-\psi\left(\frac{d}{2}\right)-\gamma \right)}{486 N^3} \biggr) \\
    & \qquad - \left(\frac{2^{1-d}\pi^{-d/2}}{\Gamma\left(\frac{d}{2}\right)}\right)^{-3} \varepsilon ^4 \frac{\alpha_2 (N-1) (N+2) }{8748 N} + \mathcal{O}(\varepsilon^5) \, ,
\end{split}
\end{equation}
which, upon matching with \eqref{eq:F3-H(N)} and \eqref{eq:F2-H(N)}, gives 
\begin{equation}
    \tilde{F}^{H_N}_{d} = \Omega(d) \ A^{H_N}_d(\alpha_2(\alpha_0)) \, ,
\end{equation}
with the same relations given in Eq.~\eqref{eq:matched-coefficients}.

Having shown that our matching of $A$ with $F$ holds at two different fixed points we may argue that this is true at every fixed point. Another way to say this is that in their tensorial versions both $A$ and $F$ present the same contractions of the coupling tensors, which behave in the same way upon projection onto a fixed point. For a matching of $A$ and $F$ to be independent of the particular fixed point, we therefore require the coefficients that we match in the two quantities to be relative to the same tensor contraction in both $A$ and in $F$. In this sense, since projection onto a fixed point only acts on the tensor contractions and not on their coefficients, it must be true that our matching is completely general. A version of this argument was first presented in Ref.~\cite{Pannell:2024sia} for the short range case.

We finally substitute \eqref{eq:matched-coefficients} in the diagonalized $C_{\alpha\beta}$ of Eq.~\eqref{eq:G-alpha-beta} to match it with the RG metric $G_{\alpha\beta}$:
\begin{equation}\label{eq:matching}
     G_{\alpha \beta}=
     \delta_{\alpha \beta} + \varepsilon \ \alpha_0 \left( 1 - \frac{3\gamma^\alpha}{8 \pi^{2(d-1)}} \right) \delta_{\alpha \beta} \,.
\end{equation}
If we set the final free parameter appropriately, i.e.\ $\alpha_0=\frac{8\pi^{2(d-1)}}{3}$, we find, in analogy with Ref.~\cite{Pannell:2025ixz}, that this choice gives $C_{\alpha\beta}$ in \eqref{eq:2PF-CFT}, but with rescaled operators
\begin{equation}\label{eq:real-operators}
    [O']^\alpha(x)=2^{2 \Delta_\alpha}\sqrt{\Omega(d)\left( 1+\varepsilon \ \frac{8 \pi^{2(d-1)}}{3} \right)}[O]^\alpha(x) \, . 
\end{equation}
The constant rescaling of the operators does not change the scaling properties, however it is necessary to perform it
to ensure the matching of the two metrics.

\section{Conclusions}\label{sect:conclusions}

We considered the perturbative RG flow of the multiscalar long-range $\phi^4$ theory in dimension $d<4$ and showed that it possesses a gradient structure up to the third loop order in the coupling in terms of a scalar function $A$ and a metric $G_{IJ}$ in the space of couplings. This result extends a similar analysis performed for the short-range model in Ref.~\cite{Pannell:2025ixz}. In the case of the long-range model there is the additional complication that the RG coefficients depend on the infinitesimal parameter $\varepsilon$, which is directly related to the parameter $s$ related to the strength of the long-range interaction. The metric $G_{IJ}$ is positive definite in the perturbative regime, which implies that $A$ is a monotonic quantity along RG trajectories at this order of perturbation theory.

We complemented the analysis using conformal perturbation theory on a $d$-dimensional sphere $S_d$ to compute the free-energy $\Tilde{F}$ and the renormalized two-point function $C_{IJ}$ of weakly relevant operators at any fixed point in $d=2,3$ to the leading nontrivial order. 
These quantities have been argued to give explicit CFT candidates for the tensors of the gradient structure, and, in fact, we show that it is possible to match consistently the RG results with the CFT ones after appropriately fixing the parameters that are left undetermined in the RG computation. For the matching
we have used RG fixed points with $O(N)$ and $H_N$ symmetries, but we believe that it should be possible in general. Another interesting fact is that the matching of the Zamolodchikov's metric from CFT with the RG counterpart requires an appropriate constant rescaling of the scaling operators, similarly to what was observed in the short-range case \cite{Pannell:2024sia}.

In practice our work shows explicitly that, for some CFTs that do not admit a local energy-momentum tensor (i.e.\ the long-range models), a strong monotonicity theorem still holds perturbatively in $d<4$, and that the monotonic function can be identified with the analytical continuation of the sphere free-energy in agreement with the expectations of the $F$-theorem.
This results corroborates our Wilsonian intuition that monotonicity theorems should be rather general statements (at least weakly), because the RG comes from the averaging process in physical systems and naturally identifies a direction between UV and IR.

It is less obvious whether our result should hold beyond the third order in perturbation theory. In the case of short-range interactions, starting from five loops, a monotonic $A$ function comes from integrating $B$-functions rather than $\beta$-functions \cite{Pannell:2024sia}.
To clarify, $B$-functions are generalizations of the $\beta$-functions that incorporate an antisymmetric gamma-like term that can be interpreted as a scale-dependent flavor rotation \cite{Jack:1990eb,Herren:2021yur} (importantly, in even dimensions we have that $B=0$ implies conformal invariance because $\langle T\rangle \sim B^I O_I$, while $\beta=0$ implies only scale invariance at least for short-range models).
Terms coming from a scale-dependent flavor rotation are allowed also in the long-range model at higher loops
(and in principle they could include all the same tensor structures as the short-range one), but testing if they are needed in the long-range case would require two additional loop orders beyond the currently known result.

A strong motivation to pursue further this computation could come from establishing unequivocally that the long-range models are CFTs. It would also be interesting to find useful necessary/sufficient conditions to conformal invariance, similar to the expectation value of the trace of the energy-momentum tensor, that work for the long-range case and correctly identify the type of RG structure that should be considered in general (in analogy with $\langle T\rangle \sim B^I O_I$ in even dimensions).

Nonunitary theories have been previously argued to violate the generalized $F$-theorem, but some form of monotonicity theorem is still expected to hold in the form of a RG-decreasing function that describes the effective degrees of freedom of the theory \cite{Fei:2015oha, Giombi:2024zrt, Diatlyk:2026oxm}. If that is the case, it would be interesting to extend our study to a general noninteger dimension, where the long-range models are probably nonunitary like the short-range counterparts \cite{Hogervorst:2014rta}. With this premise one could study the difference between some effective monotonic $F_{\rm eff}$-function and the $\tilde{F}$ we explored.

One other important aspect would be the analysis of the geometry of the space of theories in the limit $s \to s_*$ from below, where there is a transition between long- and short-range models in the infrared (see Fig.~\ref{fig:LRI}). This is a nonperturbative regime in the $\varepsilon$-expansion used in this paper, however, according to Refs.~\cite{Behan:2017dwr,Behan:2017emf} the limit admits a weakly coupled description, which may be the starting point for another RG analysis. Furthermore, the recent Ref.~\cite{Fraser-Taliente:2026iuj} argues that the short-range limit is precisely the limit in which $\tilde{F}$ is maximized as a function of $s$, which gives further motivations for the study of the free-energy in this framework.

\smallskip

\paragraph*{Note.} As this draft was being finalized,
a paper by L.~Fraser-Taliente with a some overlap with our work appeared on \emph{arXiv} \cite{Fraser-Taliente:2026iuj}.

\smallskip

\paragraph*{Acknowledgments.} We are grateful to D.~Benedetti, F.~Eustachon, I.~Jack and H.~Osborn for useful conversations during the development of this work, and to S.~Rychkov, A.~Abdesselam and L.~Fraser-Taliente for corrections on the draft. LB would also like to thank I.~Klebanov for clarifications on previous results regarding the $F$-function and C.~Behan for helpful correspondence on Mellin-Barnes integrals and the \texttt{MB.m} package.

\appendix

\section{Integrals on the sphere}\label{sect:MB}

The aim of this section is to illustrate the procedure to compute the Feynman diagrams over the sphere given in Sections \ref{sect:F-function} and \ref{sect:two-point}.
Useful resources for these computations are \cite{Giombi:2014xxa,Fei:2015oha,Giombi:2015haa,Behan:2025ydd,Smirnov:2012gma,Dubovyk:2022obc} and this appendix simply summarizes the key steps with some examples.

First, we recall the expression for the scalar propagators from \eqref{eq:propagator}
\begin{equation}\label{eq:propagator-Appendix}
  \begin{tikzpicture}[scale=0.5,baseline=(vert_cent.base)]
	\node (vert_cent) at (0,0) {$\phantom{\cdot}$};
	\node (center) at (0,0) {};
	\def\radius{1cm};
    \node[inner sep=0pt] (tl) at (180:\radius) {};
    \node[inner sep=0pt] (tl1) at (180:1.5\radius) {$x$};
	\node[inner sep=0pt] (tr) at (0:\radius) {};
    \node[inner sep=0pt] (tr1) at (0:1.5\radius) {$y$};
        \filldraw (tl) circle[radius=2pt];
        \filldraw (tr) circle[radius=2pt];
	\draw (tr) -- (tl);
	\end{tikzpicture}
 \, 
 =  \frac{C_{d,s}}{\sigma(x,y)^{d-s}} \,,
 \qquad C_{d,s}=\frac{\Gamma(\frac{d-s}{2})}{\pi^{\frac{d}{2}}2^s\Gamma(\frac{s}{2})} \, ,
\end{equation}
that we can use to express every diagram in terms of coordinate-space integrals.
The function $\sigma(x,y)$ is the chordal distance between the points with coordinates $x$ and $y$ on the sphere of radius $R$ (not to be confused with the curvature scalar).

If we use stereographical coordinates, the chordal distance and the invariant integration measure become
\begin{equation}
    \sigma(x,y)=\frac{2R \ |x-y|}{\sqrt{(1+x^2)}\sqrt{(1+y^2)}} \, , \qquad \int {\rm d}^dx \sqrt{g_x} = \int {\rm d}^dx \frac{(2R)^d}{(1+x^2)^d} \, .
\end{equation}
Now let $z$ be the coordinate of any point of integration, it is always possible to rotate the other points in such a way to cancel $z$ in the integrand and carry out the $z$ integration trivially
\begin{equation*}
    \int {\rm d}^dz \sqrt{g_z} \left( \cdots\right)=\text{Vol}\left( S^d \right) \left( \cdots\right)_{z=0}
    \equiv \frac{2 \pi^{\frac{d+1}{2}}R^d}{\Gamma\left( \frac{d+1}{2} \right)}\left( \cdots\right)_{z=0} \,.
\end{equation*}
Given that $\sqrt{g}|_{z=0} =(2R)^d$ this operation is equivalent to setting $z=0$ everywhere
before integrating (including in its measure) and replacing its integration with a prefactor, i.e., $\int {\rm d}^dz \to \frac{2^{1-d} \pi^{\frac{d+1}{2}}}{\Gamma\left( \frac{d+1}{2} \right)}$.
Using this simplification, the scalar integrals over $n$-points given in Section \ref{sect:SphereFreeEnergy} have the following form
\begin{equation}
    I_n=(2R)^{n(d-2\alpha)} \frac{2^{1-d} \pi^{\frac{d+1}{2}}}{\Gamma\left( \frac{d+1}{2} \right)} \int \prod_{i=1}^{n-1} \frac{{\rm d}^d x_i}{(a+x_i^2)^{d-2\alpha}}\frac{1}{\prod_{a<b}^{n-1}|x_a-x_b|^{2\beta_{ab}}} \ ,
\end{equation}
with $\alpha=d-s$ being the power in the propagator and $\beta_{ab}$ being the power of the relative term in the denominator. Similar manipulations can be performed on the diagrams of Section \ref{sect:two-point}.

The simplified integrals enjoy a new diagrammatical form, introduced
in Ref.~\cite{Fei:2015oha}, in which we represent stereographic factors coming from measures and chordal distances as
\begin{equation}\label{eq:graphic-prescriptions}
    \begin{tikzpicture}[scale=0.5,baseline=(vert_cent.base)]
	\node (vert_cent) at (0,0) {$\phantom{\cdot}$};
	\node (center) at (0,0) {};
	\def\radius{1cm};
    \node[inner sep=0pt] (tl) at (180:\radius) {};
    \node[inner sep=0pt] (tl1) at (90:0.5\radius) {$a$};
	\node[inner sep=0pt] (tr) at (0:\radius) {};
    \node[inner sep=0pt] (tr1) at (0:1.5\radius) {$x$};
        \filldraw (tr) circle[radius=2pt];
	\draw[dashed] (tr) -- (tl);
	\end{tikzpicture} = \frac{1}{(1+x^2)^a} \ , \quad \begin{tikzpicture}[scale=0.5,baseline=(vert_cent.base)]
	\node (vert_cent) at (0,0) {$\phantom{\cdot}$};
	\node (center) at (0,0) {};
	\def\radius{1cm};
    \node[inner sep=0pt] (tl) at (180:\radius) {};
    \node[inner sep=0pt] (tl1) at (180:1.5\radius) {$x$};
    \node[inner sep=0pt] (tl2) at (90:0.5\radius) {$b$};
	\node[inner sep=0pt] (tr) at (0:\radius) {};
    \node[inner sep=0pt] (tr1) at (0:1.5\radius) {$y$};
    \filldraw (tl) circle[radius=2pt];
    \filldraw (tr) circle[radius=2pt];
	\draw (tr) -- (tl);
	\end{tikzpicture} = \frac{1}{|x-y|^{2b}} \ . 
\end{equation}
The diagrammatic expressions can then be simplified using the Mellin-Barnes representation and the rules given in Ref.~\cite[Appendix B]{Fei:2015oha}.
In order to make this Appendix self-contained we will be reporting the ones that will be of use to us here.
\begin{equation}\label{eq:MB-rules}
    \begin{split}
        &\int {\rm d}^d x \ {\rm d}^dy \quad \begin{tikzpicture}[scale=0.5,baseline=(vert_cent.base)]
	\node (vert_cent) at (0,0) {$\phantom{\cdot}$};
	\node (center) at (0,0) {};
	\def\radius{1cm};
    \node[inner sep=0pt] (l) at (180:1.5\radius) {};
    \node[inner sep=0pt] (l2) at (180:4.0\radius) {};
    \node[inner sep=0pt] (l1) at (160:1.8\radius) {$x$};
    \node[inner sep=0pt] (l12) at (170:3.0\radius) {$a_1$};
    \node[inner sep=0pt] (l13) at (90:0.5\radius) {$b$};
	\node[inner sep=0pt] (r) at (0:1.5\radius) {};
    \node[inner sep=0pt] (r2) at (0:4.0\radius) {};
    \node[inner sep=0pt] (r1) at (20:1.8\radius) {$y$};
    \node[inner sep=0pt] (r12) at (10:3.0\radius) {$a_2$};
    \node[inner sep=0pt] (tr) at (45:3.0\radius) {};
        \filldraw (l) circle[radius=2pt];
        \filldraw (r) circle[radius=2pt];
	\draw (r) -- (l);
    \draw[dashed] (l2) -- (l);
    \draw[dashed] (r) -- (r2);
	\end{tikzpicture} = \Gamma_0 (a_1,a_2,b) \\
    & \int {\rm d}^dx  \quad \begin{tikzpicture}[scale=0.5,baseline=(vert_cent.base)]
	\node (vert_cent) at (0,0) {$\phantom{\cdot}$};
	\node (center) at (0,0) {};
	\def\radius{1cm};
    \node[inner sep=0pt] (l) at (180:1.5\radius) {};
    \node[inner sep=0pt] (l2) at (180:4.0\radius) {};
    \node[inner sep=0pt] (l1) at (160:1.8\radius) {$x$};
    \node[inner sep=0pt] (l12) at (170:3.0\radius) {$a$};
    \node[inner sep=0pt] (l13) at (90:0.5\radius) {$b$};
	\node[inner sep=0pt] (r) at (0:1.5\radius) {};
    \node[inner sep=0pt] (r1) at (20:1.8\radius) {$y$};
        \filldraw (l) circle[radius=2pt];
        \filldraw (r) circle[radius=2pt];
	\draw (r) -- (l);
    \draw[dashed] (l2) -- (l);
	\end{tikzpicture} = \frac{1}{2 \pi i}\int_{-i \infty}^{i \infty} {\rm d}z_1 \,  \Gamma_1(a,b|z_1) \quad \begin{tikzpicture}[scale=0.5,baseline=(vert_cent.base)]
	\node (vert_cent) at (0,0) {$\phantom{\cdot}$};
	\node (center) at (0,0) {};
	\def\radius{1cm};
    \node[inner sep=0pt] (l) at (180:1.5\radius) {};
    \node[inner sep=0pt] (l2) at (180:4.0\radius) {};
    \node[inner sep=0pt] (l1) at (180:0.5\radius) {$y$};
    \node[inner sep=0pt] (l12) at (170:3.0\radius) {$-z_1$};
        \filldraw (l) circle[radius=2pt];
    \draw[dashed] (l2) -- (l);
	\end{tikzpicture} \\
     &\int {\rm d}^d x \quad \begin{tikzpicture}[scale=0.5,baseline=(vert_cent.base)]
    \node (vert_cent) at (0,0) {$\phantom{\cdot}$};
	\node[inner sep=0pt] (center) at (0,0) {};
	\def\radius{1cm};
    \node[inner sep=0pt] (tr) at (45:1.5\radius) {};
    \node[inner sep=0pt] (tr1) at (45:2.0\radius) {$y$};
    \node[inner sep=0pt] (br) at (315:1.5\radius) {};
    \node[inner sep=0pt] (br1) at (315:2.0\radius) {$z$};
    \node[inner sep=0pt] (l) at (180:2.0\radius) {};
    \node[inner sep=0pt] (l1) at (160:1.0\radius) {$a$};
    \node[inner sep=0pt] (cc) at (260:0.5\radius) {$x$};
    \node[inner sep=0pt] (tmp1) at (80:0.9\radius) {$b_1$};
    \node[inner sep=0pt] (tmp2) at (345:0.9\radius) {$b_2$};
        \filldraw (center) circle[radius=2pt];
        \filldraw (tr) circle[radius=2pt];
        \filldraw (br) circle[radius=2pt];
	\draw[dashed] (l) -- (center);
    \draw (center) -- (tr);
    \draw (center) -- (br);
	\end{tikzpicture} = \frac{1}{(2\pi i)^3}\int_{-i\infty}^{i\infty} {\rm d}z_1 \ {\rm d}z_2 \ {\rm d}z_3 \ \Gamma_2(a,b_1,b_2|z_1,z_2,z_3) \begin{tikzpicture}[scale=0.5,baseline=(vert_cent.base)]
	\node (vert_cent) at (0,0) {$\phantom{\cdot}$};
	\node (center) at (0,0) {};
	\def\radius{1cm};
    \node[inner sep=0pt] (l) at (180:1.5\radius) {};
    \node[inner sep=0pt] (l2) at (180:4.0\radius) {};
    \node[inner sep=0pt] (l1) at (165:1.8\radius) {$y$};
    \node[inner sep=0pt] (l12) at (170:3.0\radius) {$-z_1$};
    \node[inner sep=0pt] (l13) at (90:0.5\radius) {$-z_3$};
	\node[inner sep=0pt] (r) at (0:1.5\radius) {};
    \node[inner sep=0pt] (r2) at (0:4.0\radius) {};
    \node[inner sep=0pt] (r1) at (20:1.8\radius) {$z$};
    \node[inner sep=0pt] (r12) at (10:3.0\radius) {$-z_2$};
    \node[inner sep=0pt] (tr) at (45:3.0\radius) {};
        \filldraw (l) circle[radius=2pt];
        \filldraw (r) circle[radius=2pt];
	\draw (r) -- (l);
    \draw[dashed] (l2) -- (l);
    \draw[dashed] (r) -- (r2);
	\end{tikzpicture}
    \end{split}
\end{equation}
where
\begin{equation}\label{eq:Gammas}
    \begin{split}
        & \Gamma_0(a_1,a_2,b)=\frac{\pi^d}{\Gamma(d/2)}\frac{\Gamma(d/2-b)\Gamma(a_1+b-d/2)\Gamma(a_2+b-d/2)\Gamma(a_1+a_2+b-d)}{\Gamma(a_1)\Gamma(a_2)\Gamma(a_1+a_2+2b-d)} \ , \\
        & \Gamma_1 (a,b|z_1)=\frac{\pi^{d/2}\Gamma(-z_1)\Gamma(a+b-d/2+z_1)\Gamma(b+z_1)\Gamma(d-a-2b-z_1)}{\Gamma(a)\Gamma(b)\Gamma(d-a-b)} \ , \\
        & \Gamma_2(a,b_1,b_2|z_1,z_2,z_3)=\frac{\pi^{d/2}\prod_{i=1}^3 \Gamma(-z_i)\Gamma(a+b_1+b_2-d/2+\sum_{i=1}^3z_i)}{\Gamma(a)\Gamma(b_1)\Gamma(b_2)\Gamma(d-a-b_1-b_2)} \times \\
        & \qquad \qquad \times \Gamma(b_1+z_1+z_3)\Gamma(b_2+z_2+z_3)\Gamma(d-a-2b_1-2b_2-z_1-z_2-2z_3) \ .
    \end{split}
\end{equation}
The final result is a (potentially complicate) Mellin-Barnes integral that we need to expand in orders of $\varepsilon$.
This can be done using the Mathematica package \texttt{MB.m} \cite{Czakon:2005rk}\footnote{We recommend Ref.~\cite[Appendix D]{Behan:2025ydd} for a crash course.}.
We give an example of the above procedure in the next subsection.

\subsection{Computing an example}\label{sect:MB-example}

As an example of the procedure outlined above and to keep some promises made in Section \ref{sect:2loop-F}, we will now compute the integral from \eqref{eq:A2Fb}.
Ignoring the $\mathcal{N}$ prefactor, our diagram can be written using the scalar propagators of \eqref{eq:propagator-Appendix} as explained above, to obtain
\begin{equation*}
    \begin{tikzpicture}[scale=0.5,baseline=(vert_cent.base)]
	\node (vert_cent) at (0,0) {$\phantom{\cdot}$};
	\node (center) at (0,0) {};
	\def\radius{1cm};
	\node[inner sep=0pt] (tr) at (45:\radius) {};
	\node[inner sep=0pt] (tl) at (135:\radius) {};
	\node[inner sep=0pt] (bl) at (225:\radius) {};
    \node[inner sep=0pt] (br) at (315:\radius) {};
    \node[inner sep=0pt] (tr1) at (45:1.5\radius) {$x$};
	\node[inner sep=0pt] (tl1) at (135:1.5\radius) {$y$};
	\node[inner sep=0pt] (bl1) at (225:1.5\radius) {$z$};
    \node[inner sep=0pt] (br1) at (315:1.5\radius) {$w$};
	\draw (center) circle[radius=\radius];
	\filldraw (tr) circle[radius=2pt];
        \filldraw (tl) circle[radius=2pt];
        \filldraw (bl) circle[radius=2pt];
        \filldraw (br) circle[radius=2pt];
	\draw (tr) to[out=225, in=135] (br);
	\draw (tl) to[out=-45, in=45] (bl);
  \draw (tl) to[out=-45, in=225] (tr);
 \draw (bl) to[out=45, in=135] (br);
    \end{tikzpicture} = C_{d,s}^8 \int {\rm d}^dx \sqrt{g_x}\int {\rm d}^dy \sqrt{g_y}\int {\rm d}^dz \sqrt{g_z}\int {\rm d}^dw  \sqrt{g_w} \, \frac{1}{[\sigma(x,y)\sigma(y,z)\sigma(z,w)\sigma(w,x)]^{2\alpha}} \ ,
\end{equation*}
where $\alpha=d-s$. Moving to stereographical coordinates, we use rotational invariance to eliminate $w$ from the integrand and integrate over it, obtaining
\begin{equation}\label{eq:stereo-0}
     C_{d,s}^8 (2R)^{4(d-2\alpha)} \frac{2^{1-d} \pi^{\frac{d+1}{2}}}{\Gamma\left( \frac{d+1}{2} \right)} \int {\rm d}^dx \int {\rm d}^dy \int {\rm d}^dz \ \frac{\left[(1+x^2)(1+y^2)(1+z^2)\right]^{2\alpha-d}}{|x-y|^{2\alpha}|y-z|^{2\alpha}} \, .
\end{equation}
Ignoring the prefactors again, we turn our attention to the remaining integrals, expressing them with the graphical prescriptions from \eqref{eq:graphic-prescriptions},
\begin{equation}\label{eq:stereo}
     \int {\rm d}^dx \int {\rm d}^dy \int {\rm d}^dz \quad 
     \begin{tikzpicture}[scale=0.5,baseline=(vert_cent.base)]
	\node (vert_cent) at (0,0) {$\phantom{\cdot}$};
	\node (center) at (0,0) {};
	\def\radius{1cm};
    \node[inner sep=0pt] (l) at (180:3.0\radius) {};
    \node[inner sep=0pt] (l2) at (180:6.0\radius) {};
    \node[inner sep=0pt] (l1) at (170:3.0\radius) {$x$};
    \node[inner sep=0pt] (l12) at (175:5.5\radius) {$d-2\alpha$};
    \node[inner sep=0pt] (l13) at (165:1.5\radius) {$\alpha$};
	\node[inner sep=0pt] (r) at (0:3.0\radius) {};
    \node[inner sep=0pt] (c1) at (270:0.75\radius) {$y$};
    \node[inner sep=0pt] (r2) at (0:6.0\radius) {};
    \node[inner sep=0pt] (r1) at (10:3.0\radius) {$z$};
    \node[inner sep=0pt] (r12) at (5:5.5\radius) {$d-2\alpha$};
    \node[inner sep=0pt] (r13) at (15:1.5\radius) {$\alpha$};
    \node[inner sep=0pt] (tr) at (45:3.0\radius) {};
    \node[inner sep=0pt] (tr1) at (45:3.5\radius) {$d-2\alpha$};
        \filldraw (center) circle[radius=2pt];
        \filldraw (l) circle[radius=2pt];
        \filldraw (r) circle[radius=2pt];
	\draw (r) -- (l);
    \draw[dashed] (l2) -- (l);
    \draw[dashed] (r) -- (r2);
    \draw[dashed] (center) -- (tr);
	\end{tikzpicture}
     \, .
\end{equation}
Eq. \eqref{eq:stereo} can be integrated in $z$ using the rules in \eqref{eq:MB-rules}.
\begin{equation}
    \frac{1}{2 \pi i} \int {\rm d}^dx \int {\rm d}^dy \int_{-i \infty}^{+i \infty} dz_1 \ \Gamma_1(d-2\alpha,\alpha|z_1) \ 
    \begin{tikzpicture}[scale=0.5,baseline=(vert_cent.base)]
	\node (vert_cent) at (0,0) {$\phantom{\cdot}$};
	\node (center) at (0,0) {};
	\def\radius{1cm};
    \node[inner sep=0pt] (l) at (180:1.5\radius) {};
    \node[inner sep=0pt] (l2) at (180:6.0\radius) {};
    \node[inner sep=0pt] (l1) at (160:1.8\radius) {$x$};
    \node[inner sep=0pt] (l12) at (175:4.5\radius) {$d-2\alpha$};
    \node[inner sep=0pt] (l13) at (90:0.5\radius) {$\alpha$};
	\node[inner sep=0pt] (r) at (0:1.5\radius) {};
    \node[inner sep=0pt] (r2) at (0:6.0\radius) {};
    \node[inner sep=0pt] (r1) at (20:1.8\radius) {$y$};
    \node[inner sep=0pt] (r12) at (5:5.0\radius) {$d-2\alpha-z_1$};
    \node[inner sep=0pt] (tr) at (45:3.0\radius) {};
        \filldraw (l) circle[radius=2pt];
        \filldraw (r) circle[radius=2pt];
	\draw (r) -- (l);
    \draw[dashed] (l2) -- (l);
    \draw[dashed] (r) -- (r2);
	\end{tikzpicture}
\end{equation}
where $\Gamma_1$ can be read off from \eqref{eq:Gammas}.
The domain of integration of $z_1$ is a path consisting of a vertical line that runs parallel to the imaginary axis.
The integrals in $x$ and $y$ are performed once more with the rules from \eqref{eq:MB-rules}. After restoring all the prefactors that we have dropped along the way, the final result is
\begin{equation}\label{eq:integrand}
    \begin{tikzpicture}[scale=0.5,baseline=(vert_cent.base)]
	\node (vert_cent) at (0,0) {$\phantom{\cdot}$};
	\node (center) at (0,0) {};
	\def\radius{1cm};
	\node[inner sep=0pt] (tr) at (45:\radius) {};
	\node[inner sep=0pt] (tl) at (135:\radius) {};
	\node[inner sep=0pt] (bl) at (225:\radius) {};
    \node[inner sep=0pt] (br) at (315:\radius) {};
    \node[inner sep=0pt] (tr1) at (45:1.5\radius) {$x$};
	\node[inner sep=0pt] (tl1) at (135:1.5\radius) {$y$};
	\node[inner sep=0pt] (bl1) at (225:1.5\radius) {$z$};
    \node[inner sep=0pt] (br1) at (315:1.5\radius) {$w$};
	\draw (center) circle[radius=\radius];
	\filldraw (tr) circle[radius=2pt];
        \filldraw (tl) circle[radius=2pt];
        \filldraw (bl) circle[radius=2pt];
        \filldraw (br) circle[radius=2pt];
	\draw (tr) to[out=225, in=135] (br);
	\draw (tl) to[out=-45, in=45] (bl);
  \draw (tl) to[out=-45, in=225] (tr);
 \draw (bl) to[out=45, in=135] (br);
    \end{tikzpicture} = C_{d,s}^8 (2R)^{4(d-2\alpha)} \frac{2^{1-d} \pi^{\frac{d+1}{2}}}{\Gamma\left( \frac{d+1}{2} \right)} \frac{1}{2 \pi i} \int_{-i \infty}^{+i \infty} dz_1 \ \Gamma_1(d-2\alpha,\alpha|z_1) \ \Gamma_0(d-2\alpha,d-2\alpha-z_1,\alpha) \, ,
\end{equation}
with $\Gamma_0$ from \eqref{eq:Gammas}.
The integral is now set in a form that is suitable for the $\varepsilon$ expansion in general $d$. However it is not straightforward to analytically continue the final result in $d$ itself, so we perform it in $d=2$ and $d=3$ to reproduce \eqref{eq:A2Fb}. 
We show how to do it step by step in the next subsection, using the Mathematica \texttt{MB.m} package \cite{Czakon:2005rk}.

\subsection{The $\varepsilon$-expansion of Mellin-Barnes integrals}\label{sect:MB-code}

We are going to expand \eqref{eq:integrand} around $\varepsilon
\to 0$ and integrate it using a Mathematica script composed of the snippets below. A notebook that implements this script for all the diagrams in Section \ref{sect:2loop-F} can be found in the ancillary materials to this paper.

We first set the spacetime dimension in which we want to compute our integrals. 
\begin{lstlisting}[label={lst:DD}]
DD = 2;
\end{lstlisting}
In this example we have considered $d=2$, but the extension to $d=3$ is as straightforward as changing the previous line. 

The first step is to define $C_{d,s}$ from \eqref{eq:propagator-Appendix} and $\Gamma_0$, $\Gamma_1$ and $\Gamma_2$ from \eqref{eq:Gammas}.
\begin{lstlisting}[label={lst:mb-def}]
Cs[d_, s_] := Gamma[(d - s)/2]/(Pi^(d/2)*2^s*Gamma[s/2])

Gamma0[a1_, a2_, b_] := 
  ((Pi^D)/Gamma[D/2])*
   ((Gamma[D/2 - b]*Gamma[a1 + b - D/2]*
      Gamma[a2 + b - D/2]*Gamma[a1 + a2 + b - D])/
     (Gamma[a1]*Gamma[a2]*Gamma[a1 + a2 + 2*b - D]))

Gamma1[a_, b_, z1_] := 
  ((Pi^(D/2))*Gamma[-z1]*Gamma[a + b - D/2 + z1]*
     Gamma[b + z1]*Gamma[D - a - 2*b - z1])/
   (Gamma[a]*Gamma[b]*Gamma[D - a - b])

Gamma2[a_, b1_, b2_, z1_, z2_, z3_] := 
  Pi^(D/2)*Gamma[-z1]*Gamma[-z2]*Gamma[-z3]*
   Gamma[a + b1 + b2 - D/2 + z1 + z2 + z3]*
   Gamma[b1 + z1 + z3]*Gamma[b2 + z2 + z3]*
   Gamma[D - a - 2*b1 - 2*b2 - z1 - z2 - 2*z3]/
   (Gamma[a]*Gamma[b1]*Gamma[b2]*
    Gamma[D - a - b1 - b2])
\end{lstlisting}
In the above notation, the integrand of Eq.~\eqref{eq:integrand} becomes
\begin{lstlisting}[label={lst:mb-integrand}]
integrand = 
  Gamma1[D - 2*alpha, alpha, z1]*
   Gamma0[D - 2*alpha - z1, D - 2*alpha, alpha] /. 
    {alpha -> (D - s)} /. {s -> 1/2*(D + eps)} /. {D -> DD};
\end{lstlisting}
The next block of code is where the integration happens. For a full description of the functions used we refer the reader to \cite{Czakon:2005rk} and we limit ourselves to general comments.

Using \texttt{MBoptimizedRules} we algorithmically find the integration contours in the limit \texttt{eps->0} while keeping \texttt{eps} as a fixed variable during the analytic continuation. \texttt{MBcontinue} performs the analytic continuation of the integral for \texttt{eps->0} with the integration contours found by \texttt{MBoptimizedRules}. The output of this process is a list of integrals that is fed to \texttt{MBpreselect}, which erases the subset of integrals that would vanish at the first order in the \texttt{eps}-expansion. The surviving integrals are merged together if they have the same contours by \texttt{MBmerge}. The selected few are then expanded at first order around \texttt{eps=0} using \texttt{MBexpand}.

The integration process is then completed by applying the Barnes lemmas \cite{Smirnov:2012gma}. This can be done using \texttt{DoAllBarnes}, a function from the package \texttt{barnesroutines.m} that is complementary to \texttt{MB.m} \cite{Kosower:barnesroutines}. \texttt{DoAllBarnes} acts on the output of \texttt{MBmerge} and \texttt{MBexpand} by summing up all the different residues.
\begin{lstlisting}[label={lst:mb-integral}]
rules = MBoptimizedRules[integrand, eps -> 0, {}, {eps}];
cont = MBcontinue[integrand, eps -> 0, rules];
intselect = MBmerge[MBpreselect[cont, {eps, 0, 1}]];
intexp = MBmerge[MBexpand[intselect, 1, {eps, 0, 1}]];
ListBarnes = DoAllBarnes[intexp, True];
integral = Total@Flatten[{First[#]} & /@ ListBarnes];
\end{lstlisting}

Now we should reintroduce the prefactors as in Eq.~\eqref{eq:A2Fb} with the exception of the $2R$ term and expand around \texttt{eps=0}.
\begin{lstlisting}[label={lst:mb-result}]
diagram = (2^(1 - D)*Pi^(1/2*(D + 1))/Gamma[1/2*(D + 1)]*
    integral) /. {s -> 1/2*(D + eps)} /. {D -> DD};

Series[diagram, {eps, 0, 0}]
\end{lstlisting}
This finally reproduces \eqref{eq:A2Fb} exactly.

\end{document}